\documentclass[prb,twocolumn,amsmath,amssymb,aps,preprintnumbers,superscriptaddress]{revtex4-2}
\usepackage{physics}
\usepackage[normalem]{ulem}
\usepackage{braket}
\usepackage{bbold}
\usepackage{amsmath, amsfonts, amssymb}
\usepackage{graphicx}
\usepackage{float}
\usepackage{lipsum}
\usepackage{hyperref}
\usepackage[caption=false]{subfig}
\usepackage{mathtools}
\usepackage{tikz}
\usepackage{notes2bib}
\bibnotesetup{
	note-name = ,
	use-sort-key = false
}

\usepackage{setspace}

\begin{document}
	\newcommand{\titleinfo}{Bond-Order Density Wave Phases in Dimerized Extended Bose-Hubbard Models}
	\title{\titleinfo}
	
	\author{Zeki Zeybek}
	%\email{zzeybek@physnet.uni-hamburg.de}
	\affiliation{The Hamburg Centre for Ultrafast Imaging, Universit{\"a}t Hamburg\\Luruper Chaussee 149, 22761 Hamburg, Germany}
	\affiliation{Zentrum f{\"u}r Optische Quantentechnologien, Universit{\"a}t Hamburg\\Luruper Chaussee 149, 22761 Hamburg, Germany}

	\author{Peter Schmelcher}
	\affiliation{The Hamburg Centre for Ultrafast Imaging, Universit{\"a}t Hamburg\\Luruper Chaussee 149, 22761 Hamburg, Germany}
	\affiliation{Zentrum f{\"u}r Optische Quantentechnologien, Universit{\"a}t Hamburg\\Luruper Chaussee 149, 22761 Hamburg, Germany}
	
	\author{Rick Mukherjee}
	%\email{rmukherj@physnet.uni-hamburg.de}
	\affiliation{Zentrum f{\"u}r Optische Quantentechnologien, Universit{\"a}t Hamburg\\Luruper Chaussee 149, 22761 Hamburg, Germany}
	\begin{abstract}
	    The Bose-Hubbard model (BHM) has been widely explored to develop a profound understanding of the strongly correlated behavior of interacting bosons. Quantum simulators not only allow the exploration of the BHM but also extend it to models with interesting phenomena such as gapped phases with multiple orders and topological phases. In this work, an extended Bose-Hubbard model involving a dimerized one-dimensional model of long-range interacting hard-core bosons is studied. Bond-order density wave phases (BODW) are characterized in terms of their symmetry breaking and topological properties. At certain fillings, interactions combined with dimerized hoppings give rise to an emergent symmetry-breaking leading to BODW phases, which differs from the case of non-interacting models that require an explicit breaking of the symmetry. Specifically, the BODW phase at filling $\rho=1/3$ possesses no analogue in the non-interacting model in terms of its symmetry-breaking properties and the unit cell structure. Upon changing the dimerization pattern, the system realizes topologically trivial BODW phases. At filling $\rho=1/4$, on-site density modulations are shown to stabilize the topological BODW phase. Our work provides the bridge between interacting and non-interacting BODW phases and highlights the significance of long-range interactions in a dimerized lattice by showing unique BODW phases that do not exist in the non-interacting model.
	\end{abstract}
	\maketitle
	
        \section{Introduction} 
The Bose-Hubbard model (BHM) is a paradigmatic model for understanding strongly correlated materials \cite{giamarchi_quantum_2003,jaksch98,arovas,cazalilla_one_2011} and has been intensively studied from a theoretical point of view \cite{freericks_phase_1994,ejima_dynamic_2011,anders2010,elstner1999,essler_one-dimensional_2005,batrouni_supersolids_1995}. With the advent of ultracold-atom-based quantum simulators \cite{bloch_ultracold_2005, bloch_quantum_2012, lewens2007} including Rydberg atoms \cite{weimer_rydberg_2010, browaeys_many-body_2020}, the exploration of the BHM has provided valuable insights into complex phenomena arising in many-body physics \cite{greiner_quantum_2002,esslin2010,huber2007,Wilhelm_2003,zhu_topological_2013,maschler2005}. The precise control achieved in these platforms over the system parameters has facilitated probing physical scenarios that are impossible if not difficult to achieve in conventional solid-state systems. This has sparked interest in different variants of the BHM such as the extended Bose-Hubbard model (EBHM) \cite{baier_extended_2016,Trefzger_2011}. Whereas standard BHM only comprises on-site interactions between particles, certain examples of EBHM \cite{rossini_phase_2012,Mazz2006EBHM,kraus_superfluid_2020} include off-site interactions which enrich the phase diagram of the BHM by hosting supersolid phases and the Haldane insulator \cite{mishra_supersolid_2009,dalla_torre_hidden_2006,berg_rise_2008,kottmann_supersolid-superfluid_2021}.

 In a dimerized lattice, the EBHM has been shown to reveal a wealth of physical phenomena such as density-wave phases at fractional densities and symmetry-protected topological (SPT) phases \cite{roth_phase_2003,sugimoto_quantum_2019,fraxanet_topological_2022}. In particular, EBHM with dimerized hoppings and nearest-neighbor interactions has recently been shown to host phases with both the density wave and bond orders referred to as bond-order density wave (BODW) \cite{hayashi_competing_2022,zeybek_quantum_2023}. In our recent work, we studied a system of Rydberg atoms with both dipolar and van der Waals interactions on a dimerized lattice where we explored the combined effects of dimerized hopping and long-range interactions \cite{zeybek_quantum_2023}. We identified BODW phases at different fillings, and in particular, we found it for $\rho=1/3$ with no counterpart in nearest-neighbor interacting models thereby highlighting the importance of beyond nearest-neighbor processes. Thus, we have shown that having a nontrivial unit cell structure with long-range interactions can stabilize insulating states with multiple coexisting orders. These phases were previously explored in non-interacting spinless fermionic models in 1D superlattices with periodically modulated hopping amplitudes with an emphasis on studying SPT phases rather than their symmetry-breaking properties \cite{guo_kaleidoscope_2015}. Understanding topological phenomena in interacting systems has become very important \cite{noauthor_symmetry-protected_nodate, julia-farre_revealing_2022, gonzalez-cuadra_symmetry-breaking_2019,gonzalez-cuadra_intertwined_2019} since the underlying concepts have led among others to quantized transport coefficients \cite{qhall,quantconduc}, degeneracies in Bloch bands\cite{chiu_classification_2016}, and the presence of unusual edge modes \cite{noauthor_symmetry-protected_nodate, hasan2010}. Topological band theory has been very successful in explaining the topological character of states in the non-interacting case \cite{bansil_colloquium_2016,chiu_classification_2016}. Since topological band theory depends on single-particle properties, this description cannot be directly extended to systems with interactions \cite{chen_quantum_2010,bergholtz_topological_2013,salerno_interaction-induced_2020,rachel_interacting_2018,marques_multihole_2017}. However, a thorough characterization of the BODW phases of interacting systems in terms of both the symmetry-breaking and topological properties is lacking although BO phases were studied extensively \cite{grusdt_topological_2013,fraxanet_topological_2022,gonzalez-cuadra_symmetry-breaking_2019,gonzalez-cuadra_intertwined_2019}. In particular, the relationship between interactions, lattice dimerization, and topological features has not been explored extensively, and is carried out in this work by characterizing topological and symmetry-breaking properties of BODW phases in 1D at various fillings. 

\begin{figure}[t!]
		\includegraphics[width=1\columnwidth]{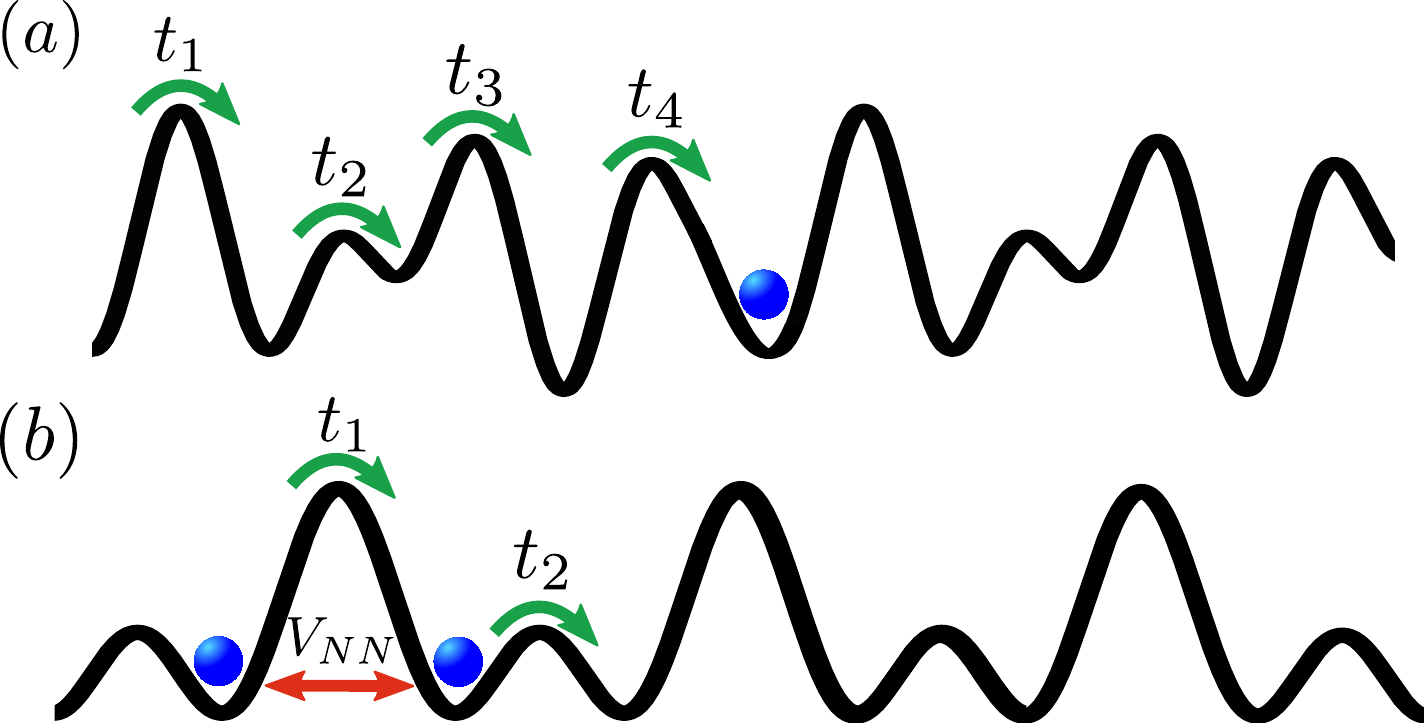}
		\caption{Superlattice structure in (a) with period $l=4$ where $t_1, t_2, t_3, t_4$ correspond to (periodic) hopping amplitudes in the non-interacting case can emerge from (b) combining repulsive nearest-neighbor interactions $V_{NN}$ in a dimerized lattice with alternating hoppings $t_1$ and $t_2$. } 
        \label{fig:slatt}
\end{figure}

 Our results indicate that dimerized hoppings combined with long-range interactions lead to BODW phases with emergent symmetry-breaking properties. This differs from their non-interacting counterparts where the symmetry-breaking pattern is provided by a superlattice with multiple hopping amplitudes [Fig.~\ref{fig:slatt}(a)]. For example, at the filling $\rho=1/4$, combining NN interactions with dimerized hopping realizes an emergent $\text{BODW}_{1/4}$ phase [Fig.~\ref{fig:slatt}(b)] whose non-interacting analog requires a tetramerized lattice with four hopping amplitudes [Fig.~\ref{fig:slatt}(a)]. By tuning dimerized hopping amplitudes, topologically trivial BODW phases can be realized. We demonstrate that a non-trivial $\text{BODW}_{1/4}$ configuration can be obtained by applying repulsive on-site pinning potentials. At filling $\rho=1/3$, a $\text{BODW}_{1/3}$ phase exhibits symmetry-breaking properties with no analog in the non-interacting case. 

\begin{figure*}[t!]
		\includegraphics[width=1\textwidth]{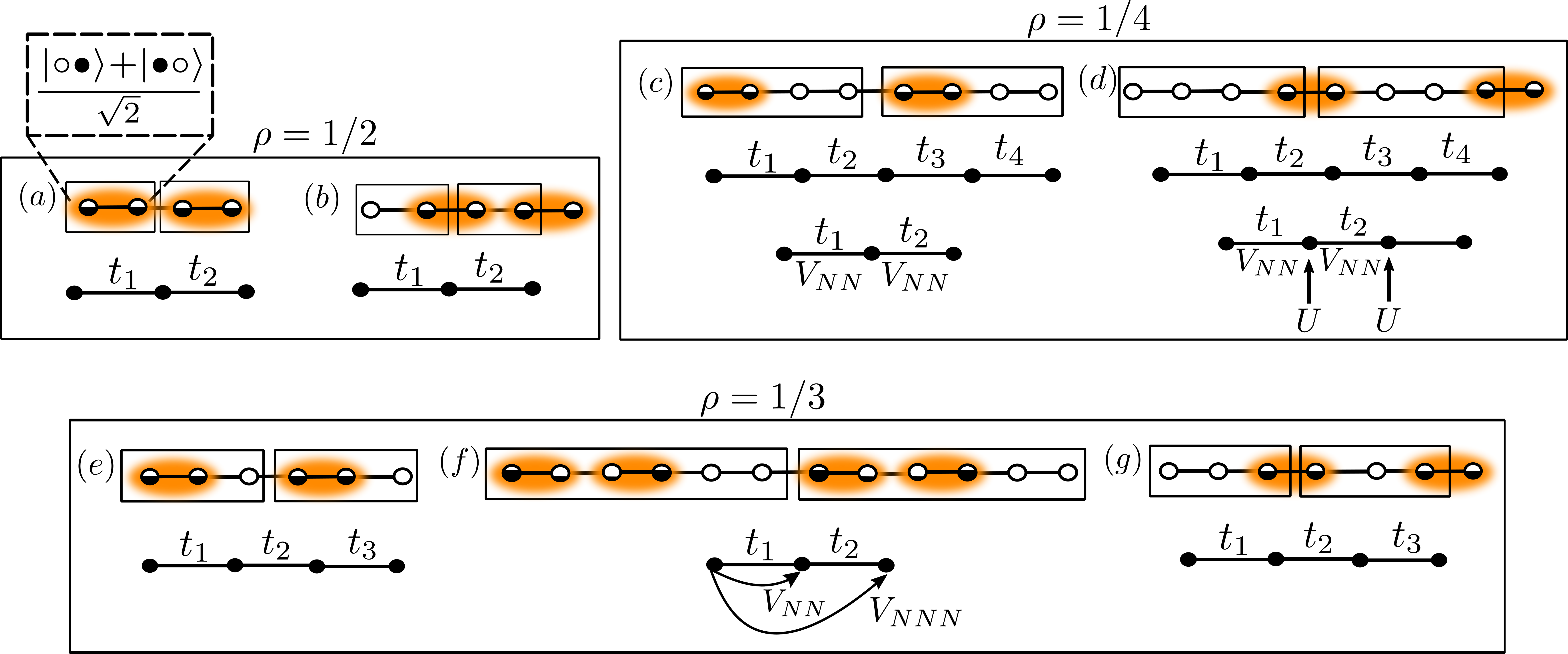}
		\caption{Schematic depictions of the BODW phases are displayed. At $\rho=1/2$, the BO phase in the (a) trivial and (b) topological sectors is shown. Unit cells are denoted by rectangles. In (a) intra-cell and (b) inter-cell dimers are formed when $|t_1|>|t_2|$ and $|t_1|<|t_2|$ respectively. Dimers are expressed by $(\ket{\circ \bullet}+\ket{\bullet \circ})/\sqrt{2}$,  where $\circ$ and $\bullet$ denote empty and boson occupied site. (c) Topologically trivial $\text{BODW}_{1/4}$ phase is shown. This phase can be found i) in the non-interacting model with hopping amplitudes $[t_1,t_2,t_3,t_4]$ and ii) in the model with NN interactions and dimerized hopping amplitudes $|t_1|>|t_2|$. (d) Topologically non-trivial $\text{BODW}_{1/4}$ with inter-cell dimers connecting two unit cells is shown. This phase can be obtained from i) the non-interacting model with hopping constraints $|t_1| = |t_3|$ and $|t_2|<|t_4|$ and ii) the NN interacting setup with repulsive on-site potentials $U$ at every second and third site inside the unit cell. $(e)$ Trivial $\text{BODW}_{1/3}$ phase in the non-interacting system with hopping amplitudes $[t_1,t_2,t_3]$ is shown. $(f)$ $\text{BODW}_{1/3}$ in the model with NN and NNN interactions with dimerized hopping $|t_1|>|t_2|$ is depicted. $(g)$ is the topological $\text{BODW}_{1/3}$ with inter-cell dimers and can be obtained in the non-interacting model with $|t_1| = |t_2|$ and $|t_1|<|t_3|$.} 	
        \label{fig:phase_diag}
\end{figure*}
\section{Hamiltonian, Methodology and Phases}
In the following, we introduce the model Hamiltonian and discuss its ground-state properties for certain limiting cases. We provide details about the employed numerical method and the observables used for characterizing the ground states and extracting corresponding topological properties.
\subsection{Hamiltonian and specific phases} \label{sec:phases}
We consider a 1D lattice of hardcore bosons with nearest-neighbor hopping and long-range repulsive density-density interactions. The EBHM with hardcore bosons in this configuration is described by 
\begin{align}\label{eqn:EBHM_Ryd_Dim}
		\hat H &= -\sum_{i} (t_{i,i+1} \hat b^\dagger_i \hat b_{i+1} + \text{H.c.}) +\sum_{i<j} V_{i,j}  \hat n_i\hat n_{j} \notag\\
  &+ U\sum_i \hat n_i,
	\end{align}
 where $\hat b_i $ $(\hat b^{\dagger}_i)$ annihilates (creates) a boson at site $i$ and $\hat n_i=\hat b^{\dagger}_i \hat b_i$ is the number operator. The tunneling amplitude $t_{i,i+1}$ is the nearest neighbor hopping amplitude between site $i$ and $i+1$. In Rydberg atom platforms, it can be encoded by dipolar exchange interactions between highly excited atoms. The off-site interaction terms $\hat n_i \hat n_j$ give an energy penalty for nearby bosons and can be implemented in Rydberg atom quantum simulators through van der Waals interactions \cite{browaeys_many-body_2020, weimer_rydberg_2010}. Due to the scaling with the inter-particle distance as $1/|i-j|^6$, the range is restricted up to next-nearest-neighbor with $|i-j|=2$ in Eq.~(\ref{eqn:EBHM_Ryd_Dim}) above and the interaction strength is tuned by $V_{i,j}$. For any given site $i$, nearest-neighbor (NN), and next-nearest-neighbor (NNN) interaction strengths are denoted by $V_{i,i+1}$, $V_{i,i+2}$ respectively. For notational convenience, we also denote hopping and interaction strengths as $t_{i,i+1}=t_i$ and $V_{i,i+1}=V_{NN}$, $V_{i,i+2}=V_{NNN}$. The third term is the on-site potential with strength $U$ and can be controlled by the detuning of the Rydberg lasers \cite{browaeys_many-body_2020}. It is set to $U=0$ unless otherwise stated. 

In the case of non-interacting bosons ($V_{NN}, V_{NNN} = 0$) with $t_i=t_{i+l}$ and $l=2$, Eq.~(\ref{eqn:EBHM_Ryd_Dim}) reduces to the bosonic version of the Su-Schrieffer-Heeger (SSH) model \cite{su_solitons_1979} which was originally formulated to describe non-interacting fermionic particles hopping on a dimerized lattice with alternating hopping amplitudes. It can be mapped to the free fermion model via the Jordan-Wigner transformation \cite{PhysRevB.73.174516}. Similarly, in the presence of periodic modulations $t_i = t_{i+l}$ with period $l$, the unit cell size is enlarged by $l$ and gapped phases can be found at the filling $n/l$ with $n=1, \dots l-1$ \cite{guo_kaleidoscope_2015}. Therefore, combining periodic modulations with appropriate fillings gives rise to translational symmetry breaking in the ground state. The ground state then possesses topological properties depending on the ratios of the hopping amplitudes in the modulation pattern.

To introduce certain concepts discussed above, we start the discussion with the non-interacting model at the half-filling with dimerized hopping amplitudes $t_1/t_2 \neq 1$. In this case, the unit cell is doubled, and a symmetry-broken phase with bond order (BO) is realized \cite{grusdt_topological_2013}. The BO phase is identified as having independent dimers formed along the lattice. Dimers are given by $(\ket{\circ \bullet}+\ket{\bullet \circ})/\sqrt{2}$, where $\circ$ and $\bullet$ denote empty and boson occupied site respectively. There are two possible symmetry-broken configurations depending on whether the first link has a large ($|t_1|>|t_2|$) or small ($|t_1|<|t_2|$) hopping amplitude in the modulation pattern. $|t_1|>|t_2|$ admits a trivial phase with bond order in which dimers are formed inside the unit cell as shown in Fig.~\ref{fig:phase_diag}(a). The ground state is expressed as a product of dimers in the form $\prod\limits_{i=0}^{L-2} (\frac{\hat b^{\dagger}_{2i}+\hat b^{\dagger}_{2i+1}}{\sqrt{2}})\ket{\circ \circ \dots \circ}$. $|t_1|<|t_2|$ yields a SPT phase with BO with topologically protected edge states. In this topological sector, the dimers are of inter-cell character as depicted in Fig.~\ref{fig:phase_diag}(b), and the local Berry phase is quantized to $\pi$ at each of the inter-cell links \cite{gonzalez-cuadra_symmetry-breaking_2019,gonzalez-cuadra_intertwined_2019}. Including NN interactions ($V_{NN} \neq 0$) at half-filling results in transitioning from the BO phase to the density-wave (DW) phase \cite{hayashi_competing_2022, zeybek_quantum_2023}. Due to a finite $V_{NN}$, the system puts an energy penalty to having NN particle occupations, and dimer formation is prohibited. This leads to a DW phase described by the state $\ket{\bullet \circ \bullet \dots \bullet \circ}$ with alternating particle occupations.

The system at filling $\rho=1/4$ without interactions and $t_i=t_{i+l}$ with $l=4$ gives rise to the $\text{BODW}_{1/4}$ phase as shown in Fig.~\ref{fig:phase_diag}(c,d). Differing from the individual BO and DW phases, BODW phases exhibit breaking of the translational invariance for both the bond and site densities. By explicitly modulating the hoppings with a period of four, the unit cell is enlarged to $l=4$ and particles localize to form dimers at every fourth bond [Fig.~\ref{fig:phase_diag}(c)] and the state is described by $\prod\limits_{i=0}^{L-4} (\frac{\hat b^{\dagger}_{4i}+\hat b^{\dagger}_{4i+1}}{\sqrt{2}})\ket{\circ \circ \dots \circ}$. Different symmetry-broken sectors of the $\text{BODW}_{1/4}$ can be obtained depending on the modulation pattern similar to the $\rho=1/2$ case. Specifically, having hoppings in the form $|t_1|=|t_3|$ and $|t_2| < |t_4|$ stabilizes a topological $\text{BODW}_{1/4}$ phase protected by inversion symmetry [Fig.~\ref{fig:phase_diag}(d)]. The same phase as discussed above can be obtained in the presence of interactions with only dimerized hoppings \cite{zeybek_quantum_2023}. Interactions effectively induce superlattice structures with higher periods despite having a dimerized lattice as shown in Fig.~\ref{fig:slatt}(a,b). The model in Eq.~(\ref{eqn:EBHM_Ryd_Dim}) at the filling $\rho=1/4$ with only nearest-neighbor interactions ($V_{NNN} = 0$) and dimerized hoppings $t_i=t_{i+l}$ with $l=2$ and $|t_1|>|t_2|$ hosts the $\text{BODW}_{1/4}$ phase [Fig.~\ref{fig:phase_diag}(c)]. A dimerized lattice at $\rho=1/4$ without interactions leads to a vanishing energy costs for particle-hole excitations since the particle density $\rho=1/4$ is too dilute for such a setup. This leads to a vanishing particle-hole excitation gap where the Luttinger liquid phase is realized. Turning on NN interactions causes particle-hole excitations to come with an energy penalty due to the suppression of NN boson occupation. The ground state then energetically favors a configuration consisting of a single dimer in every unit cell with four sites. This leads to the translational symmetry breaking with an enlarged unit cell of $l=4$, thus, giving rise to the emergence of an effective superlattice structure with a higher period despite the dimerized lattice. As the system exhibits a gap, a topological phase transition can take place upon changing the hopping dimerization from $|t_1|>|t_2|$ to $|t_1|<|t_2|$ in the framework of SSH physics. However, changing the dimerization pattern gives rise to another trivial sector of the $\text{BODW}_{1/4}$ phase which differs from [Fig.~\ref{fig:phase_diag}(c)] by one lattice translation where the dimer is located at the second link instead of the first. To favor the topological $\text{BODW}_{1/4}$, repulsive on-site modulation at every second and third site inside the unit cell can be applied as shown in Fig.~\ref{fig:phase_diag}(d). As in the $\rho=1/2$ case, including beyond NN interactions leads to the DW phase described by the state of the form $\ket{\bullet \circ \circ \circ \bullet \dots \bullet \circ \circ \circ}$. Analogously to the $\text{BODW}_{1/4}$ phase, dimerized hoppings with long-range interactions (beyond NNN) can effectively realize the superlattice structure with period $l=6$, which in the non-interacting case requires $t_i=t_{i+6}$. The state is described by $\prod\limits_{i=0}^{L-6} (\frac{\hat b^{\dagger}_{6i}+\hat b^{\dagger}_{6i+1}}{\sqrt{2}})\ket{\circ \circ \dots \circ}$.

The system at filling $\rho=1/3$ without interactions and $t_i=t_{i+l}$ with $l=3$ hosts the $\text{BODW}_{1/3}$ phase. The combined effect of a periodic hopping amplitude with the filling $\rho=1/3$ enlarges the unit cell by $l=3$. Particles localize to form dimers at every third bond [Fig.~\ref{fig:phase_diag}(e)] and a state in the form is $\prod\limits_{i=0}^{L-3} (\frac{\hat b^{\dagger}_{3i}+\hat b^{\dagger}_{3i+1}}{\sqrt{2}})\ket{\circ \circ \dots \circ}$ obtained. Having a modulation pattern in the form $|t_1|=|t_2|$ and $|t_1| < |t_3|$ stabilizes a topological configuration of the $\text{BODW}_{1/3}$ phase [Fig.~\ref{fig:phase_diag}(g)]. Both non-interacting and interacting models at $\rho=1/4,1/6$ realize the same BODW phases, but the interacting model exhibits a unique BODW phase at $\rho=1/3$ which differs from the non-interacting one by $i)$ a symmetry-breaking with unit cell enlargement emerges which is not imposed at the Hamiltonian level $ii)$ bond densities in the unit cell do not correspond to dimers and have different numbers of sites [Fig.~\ref{fig:phase_diag}(e) and (f)]. The model in Eq.~(\ref{eqn:EBHM_Ryd_Dim}) with up second nearest-neighbor interactions ($V_{NN},V_{NNN} \neq 0$) and dimerized hoppings $t_i=t_{i+l}$ with $l=2$ hosts the $\text{BODW}_{1/3}$ phase at $\rho=1/3$ filling as shown in Fig.~\ref{fig:phase_diag}(f). Having NN and NNN interactions favor suppression of a boson occupation up to the next NN sites. This leads to a symmetry-breaking with a unit cell enlargement by $l=3$ where on-site particle densities oscillate with period-3. Due to dimerized hoppings, this enlargement further goes up to $l=6$ where finite bond densities occur at certain links between sites. As it can be seen from comparing the Figs.~\ref{fig:phase_diag}(e) and (g) with (f), the interacting $\text{BODW}_{1/3}$ has a unit cell of size $l=6$ and the bond densities do not correspond to dimers in the form $(\ket{\circ \bullet}+\ket{\bullet \circ})/\sqrt{2}$. Therefore, the state cannot be described by a product state of independent dimers as they are in the previous BODW phases. 

\subsection{Numerical Method and Observables}\label{sec:numeric}
As described in the previous section, BODW phases exhibit both bond and density wave order, which is characterized by the breaking of translational invariance concerning both bond and site density. This is manifested in the modulation of both site $\expval{\hat n_i}$ and bond $\braket{\hat B_i}$ density, where $\hat B_i =\hat b^{\dagger}\hat b_i + \text{h.c.}$ is the bond operator. The unit cell of the system is enlarged after the symmetry breaking. This ordering in gapped phases can be characterized by computing appropriate structure factors. To identify BO and DW characteristics in the ground state, we compute the following structure factors defined as the following,
\begin{align}
    \mathcal{S}_{\text{DW}} = \frac{1}{L^2}\sum_{i,j}e^{ikr}\braket{\hat n_i \hat n_j}, \\
    \mathcal{S}_{\text{BO}} = \frac{1}{L^2}\sum_{i,j}e^{ikr}\braket{\hat B_i \hat B_j},
\end{align}
where $\mathcal{S}_{\text{DW}}$ and $\mathcal{S}_{\text{BO}}$ correspond to the DW and BO structure factors where $\braket{\hat n_i \hat n_j}$ and $\braket{\hat B_i \hat B_j}$ probe site and bond density correlations respectively. $k$ is the crystal momentum and $r=\abs{i-j}$ denotes the distance between the sites $i$ and $j$ in the lattice. BO and DW orders translate into pronounced peaks at certain crystal momenta. This helps figure out the unit cell of the ordered phases, which provides information about the translational symmetry-breaking nature of the phase. Analogous to SSH physics, there are multiple symmetry-broken sectors of the BODW phases and they can be connected by an appropriate lattice translation. In finite-size systems, this translates into having symmetry-broken sectors of the BODW phases with identical bulk properties but different edges. Therefore, the previously introduced local order parameters cannot probe all the properties of the BODW phases such as topological properties that make the various symmetry-broken sectors different from each other. 

To gain insight into the global properties of the ground state, we calculate the entanglement spectrum, local Berry phase, and density distribution of the edge modes. The entanglement entropy is computed by partitioning the system and writing the ground state as
\begin{equation}\label{eq:BER}
        \left|\psi_{\mathrm{GS}}\right\rangle=\sum_n {\xi}_n\left|\psi_n\right\rangle_{\mathcal{L}} \otimes\left|\psi_n\right\rangle_{\mathcal{R}} , \quad \epsilon_n= -2 \log \left({\xi}_n\right),
\end{equation}
where $\mathcal{L}$ and $\mathcal{R}$ are the two subsystems, and $\xi_n$ are the corresponding Schmidt eigenvalues. The entanglement spectrum is defined as the set of all the Schmidt eigenvalues in the logarithmic scale $\epsilon_n = -2 \log(\xi_n)$. In 1D, it has been shown that under the preservation of their protecting symmetries, SPT phases exhibit degeneracies in their entanglement spectrum \cite{pollmann_entanglement_2010, pollmann_detection_2012}. In this way, the entanglement spectrum only consists of a group of degenerate Schmidt eigenvalues. 

We determine the local Berry phase \cite{hatsugai_quantized_2006, hatsu2}, which is a robust topological invariant that unequivocally identifies SPT phases in interacting models \cite{gonzalez-cuadra_symmetry-breaking_2019,gonzalez-cuadra_intertwined_2019, Dynamic_BHSSH_Fraxanet_2023}. For a Hamiltonian $H(\lambda)$ that depends on an external parameter $\lambda \in [\lambda_i, \lambda_f]$, an adiabatic cyclic evolution with $H(\lambda_i) = H(\lambda_f)$ can be considered. It was shown in \cite{hatsugai_quantized_2006} that as long as $H(\lambda)$ commutes with the antiunitary operator of the form $\hat{\Theta}=K \hat U$ where $K$ is complex conjugation and $\hat U$ is a unitary operator, the Berry phase \cite{MBerry} of the ground state $\ket{\psi_\lambda}$ as defined below,
\begin{equation} \label{eq:LocBerry}
    \gamma_C = i \oint_C d\lambda \braket{\psi_\lambda|\frac{\partial \psi_\lambda}{\partial \lambda}}, \mod{2\pi}
\end{equation}
is quantized to discrete values $0$ and $\pi$ upon performing a parallel transport on a closed path $C$ with $\lambda_f = \lambda_i$. We use the quantization of the Berry phase to define a topological order parameter as shown in \cite{hatsugai_quantized_2006}. A local perturbation that respects the symmetry of the SPT is introduced without closing the gap at one of the hopping strengths on a link $\braket{ij}$ connecting sites $i$ and $j$ as $t_{ij} \rightarrow e^{i\theta}t_{ij}$ \cite{hatsu2}. A closed path $C_{\braket{ij}}$ of parameters $\theta$ for the link $\braket{ij}$ is considered and the quantized Berry phase $\gamma_C$ in (\ref{eq:LocBerry}) is identified as the local order parameter at $\braket{ij}$. As long as the protecting symmetry is present in the system, this topological property cannot change unless the gap is closed. For the numerical calculation, the closed path $C$ is discretized into $N$ points $\lambda_0, \dots, \lambda_k,\dots, \lambda_N$ with $\lambda_k = t e^{i 2\pi k / N}$, $\lambda_N=\lambda_0$. Then, the discretized Berry phase is used for a given link $(i,i+1)$ defined by the lattice Berry connection \cite{hatsu3,hatsu4} as $\gamma^N_C(i,i+1)=\text{Arg}\prod^{N-1}_{k=0} \braket{\psi^U_{\lambda_k}|\psi^U_{\lambda_{k+1}}}$, $\ket{\psi^U_{\lambda_k}}=\ket{\psi_{\lambda_k}}\braket{\psi_{\lambda_k}|\phi}$, where $\ket{\psi_{\lambda_k}}$ is the ground state and $\ket{\phi}$ is a reference state. In the limit of large $N$, $\gamma^N_C(i,i+1)$ quickly approaches the local Berry phase $\gamma_C$ in (\ref{eq:LocBerry}). The quantity $\gamma^N_C(i,i+1)$ is independent of the $\ket{\phi}$ as long as the overlap $\braket{\psi_{\lambda_k}|\phi}$ is non-vanishing, but depends on $N$ and how the closed path is discretized. In this work, $N=10$ was enough to obtain converged results, and the ground state obtained without perturbation is used as the reference state $\ket{\phi}$.

Another signature of SPT phases is the existence of localized edge states in systems with boundaries. Their presence can be signaled from the real-space density distribution $\expval{\hat n_{i}(N)}$ of $N$ bosons along the lattice. As mentioned previously, Eq.~(\ref{eqn:EBHM_Ryd_Dim}) at $\rho=1/2$ with dimerized hoppings $|t_1|<|t_2|$ and vanishing interactions yields topological BO phase. This SPT phase exhibits a polarized edge population where one of the edge sites is entirely occupied ($\expval{\hat n_{i}(N)} \sim 1$) and the other vanishes ($\expval{\hat n_{i}(N)} \sim 0$) while possessing uniform distribution ($\expval{\hat n_{i}(N)} \sim 0.5$) in the bulk. In the trivial sector $(|t_1|>|t_2|)$, the occupations would be uniform for all sites. Therefore, the polarized edge population hints at the topological character of the BO phase \cite{edge_mishra, edge_misra_2}. Another way of probing the edge properties is to check the presence of many-body edge states by obtaining localized peaks or drops in the density distribution when adding one extra particle ($N+1$) above or one extra hole ($N-1$) below the filling of interest, respectively. In the topological sector at $\rho=1/2$, the density is uniform in the bulk for the two states ($\expval{\hat n_{i}(N+1)},\expval{\hat n_{i}(N-1)}\sim 0.5$) while displays localized peaks at both the edges as $\expval{\hat n_{i}(N+1)}\sim 1$ and drops as $\expval{\hat n_{i}(N-1)}\sim 0$. Therefore, as mentioned in previous works \cite{gonzalez-cuadra_symmetry-breaking_2019, frac_charge_num_Lewsn_2022}, these many-body edge states acquire a fractional particle number of $\pm 1/2$, which can be regarded as a bosonic analogue of charge fractionalization \cite{jackiw_charge_frac_1976}. In our study, we check for localized peaks under open boundary conditions when we add one extra particle above the filling of interest. To do that we compute the following quantity,
\begin{align}
    \Delta^p_i = \expval{\hat n_i (N+1)}-\expval{\hat n_i (N)}
\end{align}
for all sites $i$ on the lattice. $\Delta^p_i$ probes the existence of a many-body edge state through one extra particle \cite{zhou_exploring_2023}. For example, in the topological sector of $\rho=1/2$, $\Delta^p_i$ displays a localized peak at one of the edges as $\Delta^p_i \sim 1$ while it vanishes $\Delta^p_i \sim 0$ for the rest, which is due to having polarized edge occupations for the $N$ boson case at the edges with $\expval{\hat n(N)}=0,1$ and $\expval{\hat n(N+1)}\sim 1$ at both the edges for the $N+1$ case. Similar results are obtained at the $\rho=1/4$ filling for the interacting case in the topological sector, which will be shown in the results. Analogously, $\Delta^h_i = \expval{\hat n_i (N)}-\expval{\hat n_i (N-1)}$ can be defined to probe the existence of a many-body edge state through one extra hole.

We perform density-matrix-renormalization-group (DMRG) \cite{white_density_1992, white_density-matrix_1993, scholwork2005, SCHOLLWOCK2011} simulations to study the ground state properties of the model in Eq.~(\ref{eqn:EBHM_Ryd_Dim}). All DMRG simulations are performed by using the TeNPy library \cite{hauschild_efficient_2018}. In this work, both finite and infinite matrix product states (MPS) are used. Symmetry-breaking properties are probed in the thermodynamic limit by performing infinite DMRG (iDMRG) simulations. For the topological characterization, finite system sizes with open boundary conditions (OBC) are employed. A maximum MPS bond dimension of $\chi=150$ is considered. We set the relative energy error to be smaller than $10^{-9}$ to ensure convergence. During the truncation, Schmidt values smaller than $10^{-10}$ are discarded. 

\begin{figure}[t!]
		\includegraphics[width=1\columnwidth]{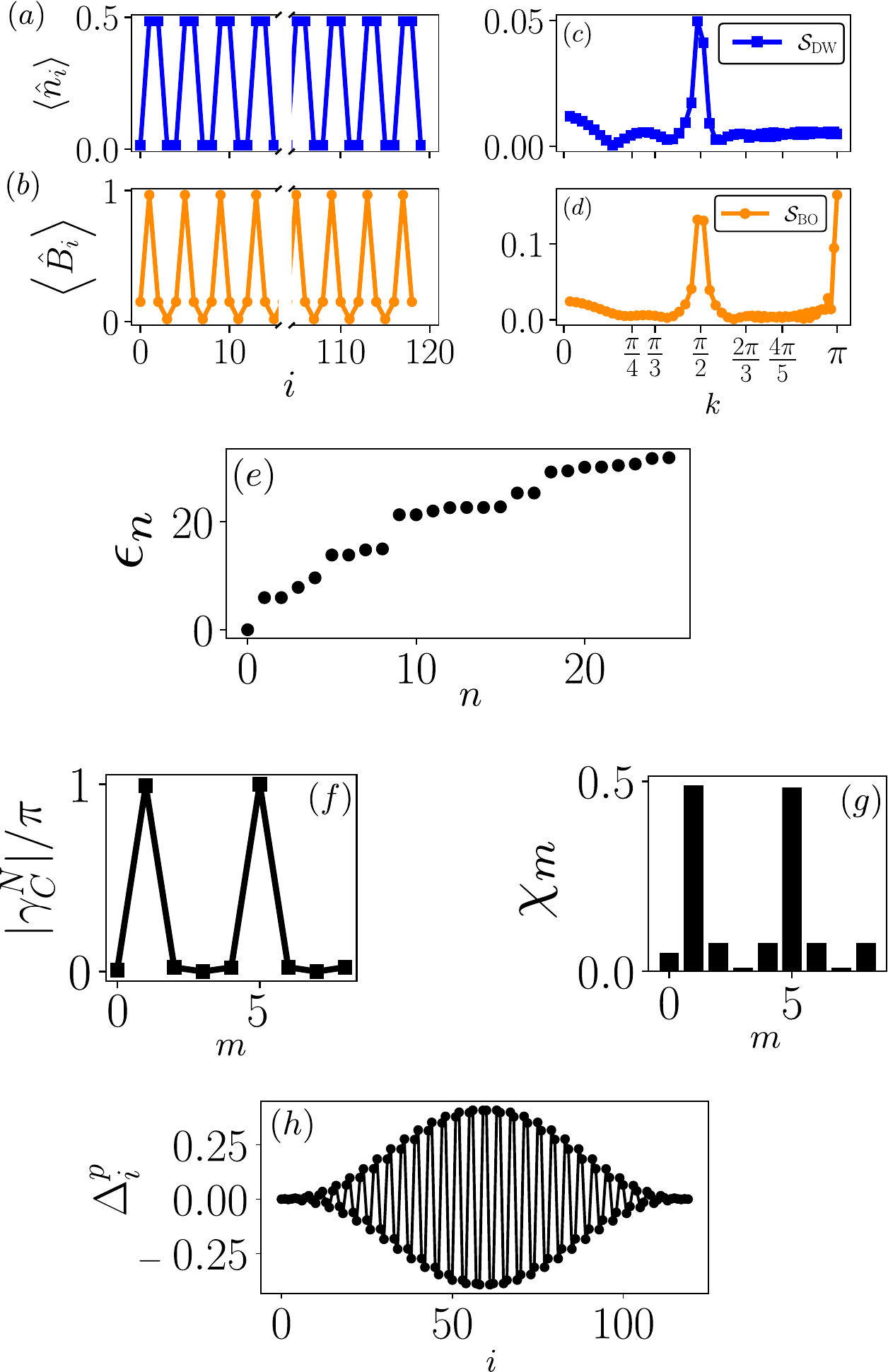}
		\caption{$\text{BODW}_{1/4}$ phase from the interacting model without on-site potential $(U=0)$ at $\rho=1/4$ in terms of symmetry-breaking (a-d) with iDMRG and topological observables (e-h) from finite DMRG simulations of the lattice size $L=120$ and OBC. Cut-out slanted lines in (a)-(b) stand for axis breaks. For the system with $t_1=0.1, t_2=1, V_{NN}=10$ [Fig.~\ref{fig:phase_diag}(c)], expectation values of the (a) site $\hat n_i$ and (b) bond $\hat B_i$ density operators are displayed. The corresponding structure factors in (c) $\mathcal{S}_{\text{DW}}$ and in (d) $\mathcal{S}_{\text{BO}}$ are shown. The entanglement spectrum $\epsilon_n$, local Berry phase $|\gamma^N_C|$, and effective hopping amplitudes $\chi_m$ for each link $m$ between two sites are shown respectively in (e,f,g) and (h) depicts $\Delta_i^p$ (see main text).} 
    \label{fig:triv_rho14}
\end{figure}

\section{Results}
In the following, we present an extensive analysis of the BODW phases in terms of their symmetry-breaking and topological characteristics. We carry out a comparative analysis where non-interacting and interacting BODW phases are contrasted to help us understand the role of interactions in giving the symmetry-breaking character. In this way, we figure out whether BODW phases in the interacting model can favor SPT configurations upon changing the dimerization pattern.  

\subsection{\texorpdfstring{$\text{BODW}_{1/4}$}{Lg}}
Figure~\ref{fig:triv_rho14} shows the symmetry-breaking and the lack of topological properties of the $\text{BODW}_{1/4}$ phase in the interacting model with $|t_1|<|t_2|$ and without on-site potentials $(U=0)$. Figs.~\ref{fig:triv_rho14}(a)-(b) depict the site and bond densities while the corresponding structure factors are provided in Figs.~\ref{fig:triv_rho14}(c)-(d). The fact that the system admits a trivial phase independent of whether $|t_1|>|t_2|$ or $|t_1|<|t_2|$ is shown by plotting the entanglement spectrum, local Berry phase, effective hoppings, and edge state population in Figs.~\ref{fig:triv_rho14}(e)-(h) respectively. The enlarged unit cell with four sites can be seen from the site and bond density profile in Fig.~(\ref{fig:triv_rho14})(a,b). In Fig.~\ref{fig:triv_rho14}(a), modulated density oscillations $\expval{\hat n_i}$ imply that a single boson is delocalized over two sites at every unit cell, thus forming a single dimer at each unit cell and giving the DW character of the phase. This differs from the non-interacting case where the particles with filling $\rho=1/4$ are loaded to the corresponding superlattice with period-4. The position of the dimer in each unit cell can be inferred from the bond densities at the links as shown in Fig.~\ref{fig:triv_rho14}(b) where a finite bond density is shown at every second link in each unit cell forming intra-cell dimers. Changing the dimerization pattern from $|t_1|<|t_2|$ to $|t_1|>|t_2|$ realizes another trivial sector of the $\text{BODW}_{1/4}$ phase with a dimer at every first link in each unit cell [Fig.~\ref{fig:phase_diag}(c)], differing from the $|t_1|<|t_2|$ case by a lattice translation. These findings also translate into the peaks of the structure factors $\mathcal{S}_{DW}(k)$ at $k=\pi/2$ and $\mathcal{S}_{BO}(k)$ at $k=\pi/2,\pi$ as shown in Fig.~\ref{fig:triv_rho14}(c,d). Since topological features are absent in this trivial phase, the entanglement spectrum does not consist of groups of degenerate Schmidt coefficients as shown in Fig.~\ref{fig:triv_rho14}(e). As mentioned in section \ref{sec:phases}, the system in the non-interacting limit at $\rho=1/4$ with periodic modulation of the hopping amplitudes $[t_1,t_2,t_3,t_4]$ as $t_i=t_{i+4}$ hosts the SPT phase as long as the constraints $|t_1|=|t_3|$ and $|t_2|<|t_4|$ are satisfied. However, in the case of Figure ~\ref{fig:triv_rho14}, the combination of the hopping dimerization $|t_1|<|t_2|$ and NN interactions ($V_{NN} \neq 0$) leads to realizing \textit{effective hopping amplitudes} that violate such constraints. This can be inferred from the effective hopping amplitudes defined over the link $m$ by $ \chi_m = |\braket{\hat b^{\dagger}_m \hat b_{m+1}}|$ \cite{guo_kaleidoscope_2015}. In particular, the effective hopping amplitudes ($\chi_2,\chi_4$) shown in Fig.~\ref{fig:triv_rho14}(g) imply that $|t_2|>|t_4|$ which violates one of the constraints mentioned. Another signature for the absence of topological features is given in Fig.~\ref{fig:triv_rho14}(f) where the local Berry phase is quantized to $0$ at the inter-cell links. This implies that the dimers are formed intra-cell and the configuration is trivial. Edge mode localization is also absent as can be seen in Fig.~\ref{fig:triv_rho14}(h) where the density distribution does not exhibit localization at the boundaries. The profile in Fig.~\ref{fig:triv_rho14}(h) signals that the system with one extra particle ($N+1$) above the filling $\rho=1/4$ displays a solitonic behavior in the density $\expval{\hat n_i (N+1)}$, which is akin to the cases with one particle above the commensurate fillings $\rho$ yielding DW phases \cite{mishra_solit_prof_2009}. The results for the $\text{BODW}_{1/6}$ phase also directly follows from the $\text{BODW}_{1/4}$ phase [Fig.~\ref{fig:triv_rho14}(a)-(d)] by changing the period $l=4$ pattern to $l=6$ for which the peaks of the structure factors are $\mathcal{S}_{DW}(k)$ at $k=\pi/3, 2\pi/3$ and $\mathcal{S}_{BO}(k)$ at $k=\pi/3,2\pi/3, \pi$. Analogously, both dimerization patterns $|t_1|>|t_2|$ and $|t_1|<|t_2|$ lead to the trivial sector of the $\text{BODW}_{1/6}$. 

\begin{figure}[t!]
		\includegraphics[width=1\columnwidth]{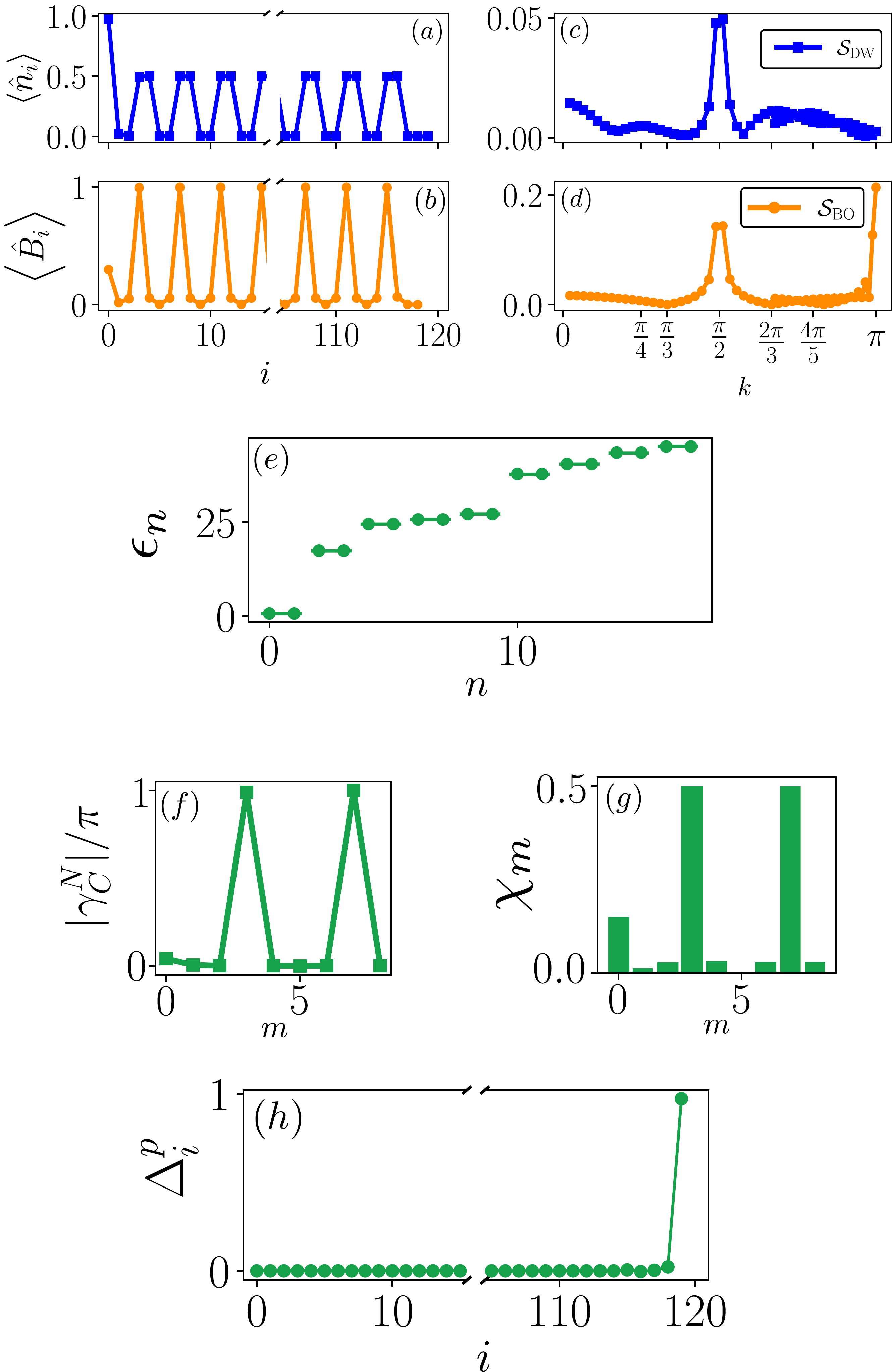}
		\caption{$\text{BODW}_{1/4}$ phase from the interacting model with on-site potential $(U \neq 0)$ at $\rho=1/4$ in terms of symmetry-breaking (a-b) with finite DMRG (c-d) with iDMRG, and topological observables (e-h) from finite DMRG simulations of the lattice size $L=120$ and OBC. Cut-out slanted lines in (a), (b) and (h) stand for axis breaks. For the system with $t_1=0.1, t_2=1, V_{NN}=10$ with on-site density modulation $U=1$ at every second and third site [Fig.~\ref{fig:phase_diag}(d)], expectation values of the (a) site $\hat n_i$ and (b) bond $\hat B_i$ density operators are displayed. The corresponding structure factors in (c) $\mathcal{S}_{\text{DW}}$ and in (d) $\mathcal{S}_{\text{BO}}$ are shown. The entanglement spectrum $\epsilon_n$, local Berry phase $|\gamma^N_C|$, and effective hopping amplitudes $\chi_m$ for each link $m$ between two sites are shown respectively in (e,f,g) and (h) depicts $\Delta_i^p$.} 
    \label{fig:spt_rho14}
\end{figure}

Figure~\ref{fig:spt_rho14} shows the symmetry-breaking and the presence of topological properties of the $\text{BODW}_{1/4}$ phase in the interacting model with $|t_1|<|t_2|$ and with on-site potentials $(U \neq 0)$. The non-trivial symmetry-broken sector of the $\text{BODW}_{1/4}$ phase with SPT is favored by adding a repulsive pinning term of the form $U \hat n_i$ at every second and third site in each unit cell to satisfy the hopping constraints [Fig.~\ref{fig:phase_diag}(d)]. As expected, the symmetry-breaking properties shown in Fig.~\ref{fig:spt_rho14}(a)-(d) are identical to the previous case without the on-site potential [Fig.~\ref{fig:triv_rho14}(a)-(d)]. However, the polarized site densities at the left edge ($\expval{\hat n_{L}}=1$) and the right edge ($\expval{\hat n_R}=0$) as shown in Fig.~\ref{fig:spt_rho14}(a) signals the existence of edge states, thus providing a signature for the existence of SPT. The entanglement spectrum exhibits twofold degeneracy with pairs of degenerate Schmidt eigenvalues as can be seen in Fig.~\ref{fig:spt_rho14}(e). This time the effective amplitudes ($\chi_2,\chi_4$) obey the constraints where $|t_2| < |t_4|$ as shown in Fig.~\ref{fig:spt_rho14}(g). This is due to having repulsive on-site density which does not favor boson occupation. This leads to the breaking of dimers at every second link in a given unit cell while promoting dimer formation at every fourth bond connecting the unit cells. This is also reflected in Fig.~\ref{fig:spt_rho14}(f) where the local Berry phase is quantized to $\pi$ at every inter-cell link. The presence of localized density of the edge mode is also shown in Fig.~\ref{fig:spt_rho14}(h), which is similar to the non-interacting topological BO phase at the filling $\rho=1/2$ mentioned in \ref{sec:numeric}.

\subsection{\texorpdfstring{$\text{BODW}_{1/3}$}{Lg}}
\begin{figure}[h!]	\includegraphics[width=1\columnwidth]{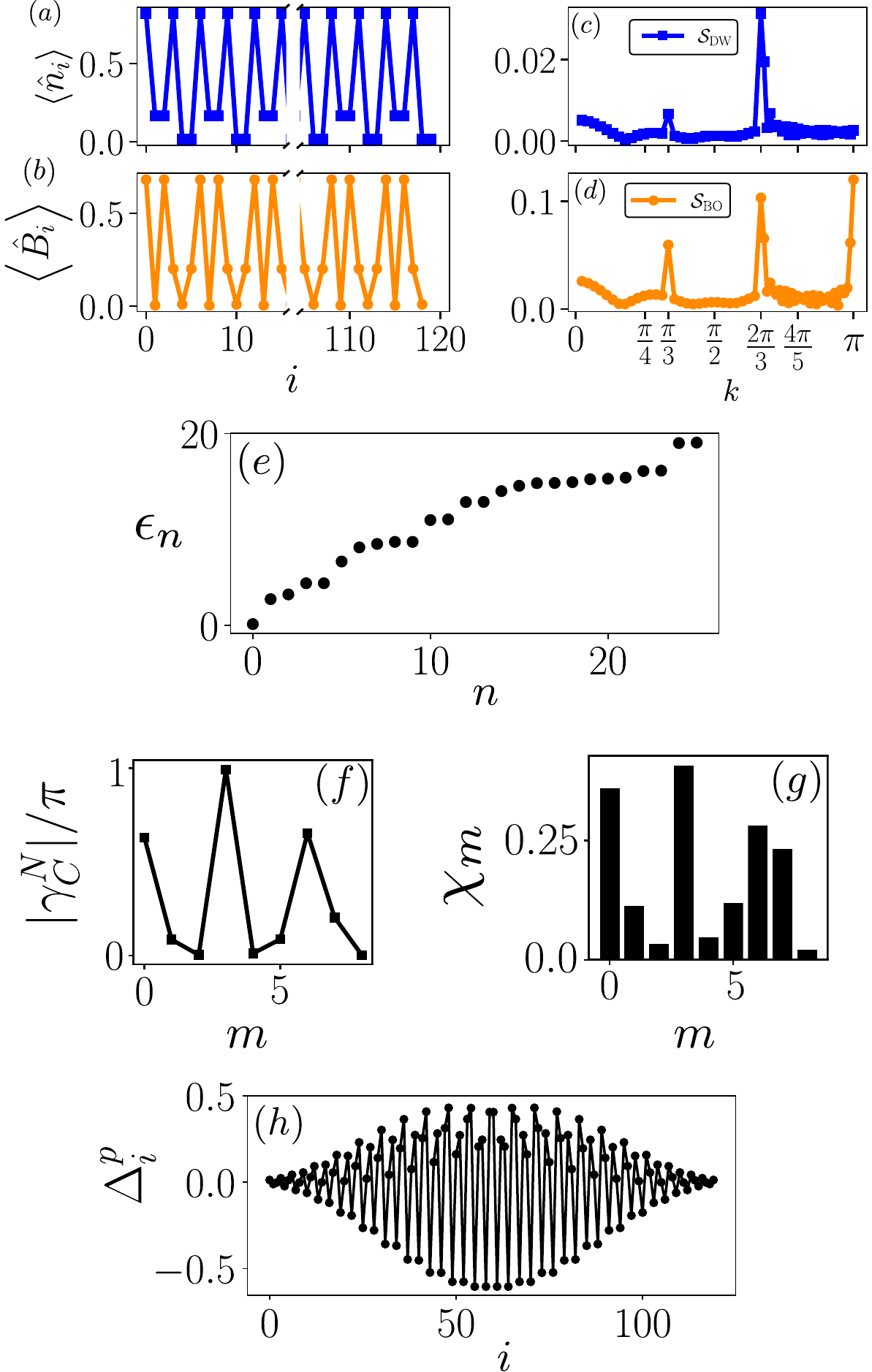}
		\caption{$\text{BODW}_{1/3}$ phase from the interacting model at $\rho=1/3$ in terms of symmetry-breaking (a-d) with iDMRG and topological observables (e-h) from finite DMRG simulations of the lattice size $L=120$ and OBC. Cut-out slanted lines in (a) and (b) stand for axis breaks. For the system with $t_1=1, t_2=0.1, V_{NN}=20, V_{NNN}=1.2$ [Fig.~\ref{fig:phase_diag}(f)], expectation values of the (a) site $\hat n_i$ and (b) bond $\hat B_i$ density operators are displayed. The corresponding structure factors in (c) $\mathcal{S}_{\text{DW}}$ and in (d) $\mathcal{S}_{\text{BO}}$ are shown. The entanglement spectrum $\epsilon_n$, local Berry phase $|\gamma^N_C|$, and effective hopping amplitudes $\chi_m$ for each link $m$ between two sites are shown respectively in (e,f,g) and (h) depicts $\Delta_i^p$.} 
          \label{fig:int_rho13}
\end{figure}

\begin{figure}[t!]
		\includegraphics[width=1\columnwidth]{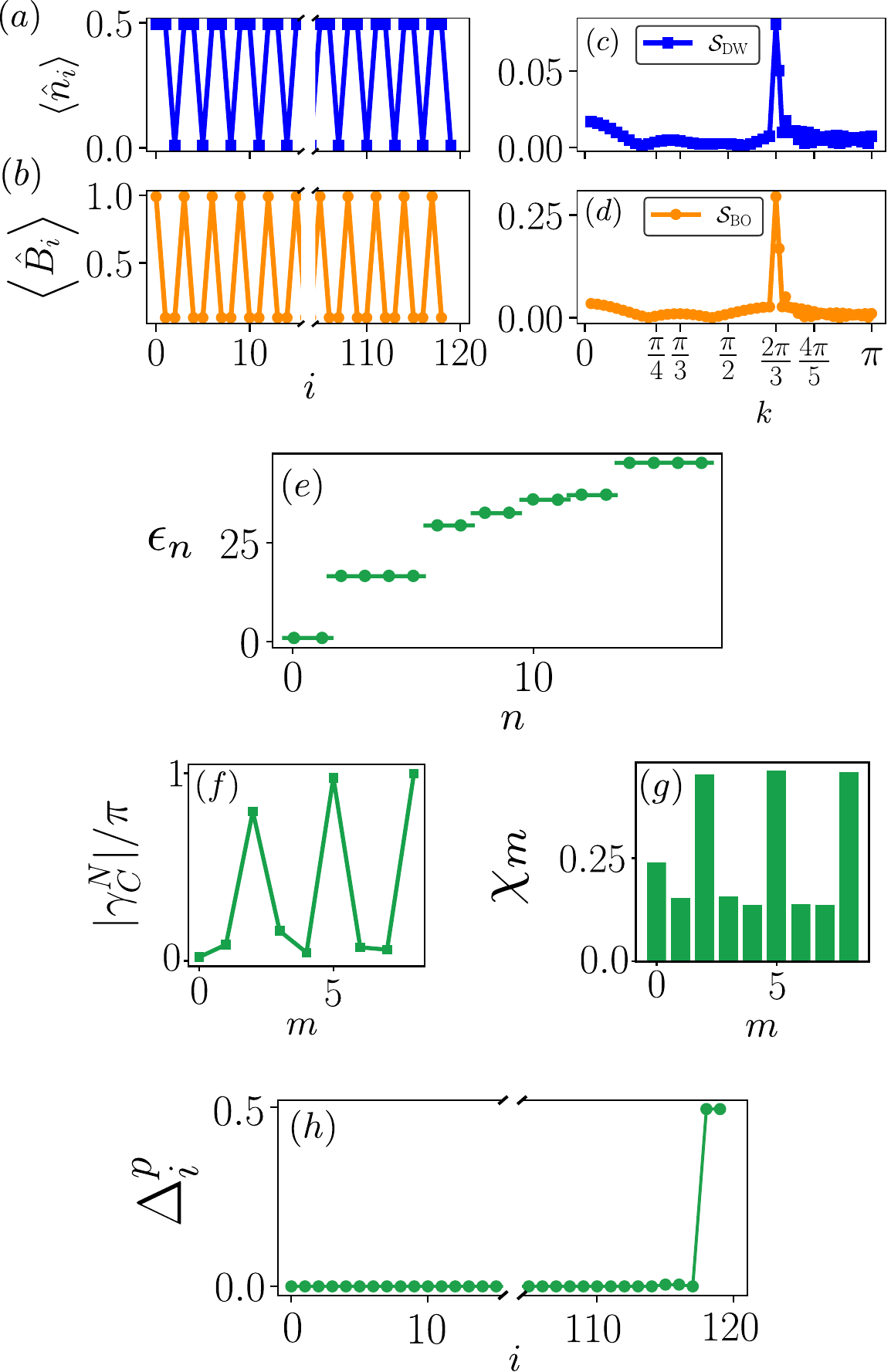}
		\caption{$\text{BODW}_{1/3}$ phase from the non-interacting model at $\rho=1/3$ in terms of symmetry-breaking (a-d) for the system with $t_1=1, t_2=0.1, t_3=0.1$ [Fig.~\ref{fig:phase_diag}(e)] with iDMRG is shown. Cut-out slanted lines in (a), (b) and (h) stand for axis breaks. Expectation values of the (a) site $\hat n_i$ and (b) bond $\hat B_i$ density operators are displayed. and the corresponding structure factors in (c) $\mathcal{S}_{\text{DW}}$ and in (d) $\mathcal{S}_{\text{BO}}$ are shown. For the system with $t_1=0.1, t_2=0.1, t_3=1$ [Fig.~\ref{fig:phase_diag}(g)], topological observables (e-h) from finite DMRG simulations of the lattice size $L=120$ and OBC are obtained. The entanglement spectrum $\epsilon_n$, local Berry phase $|\gamma^N_C|$, and effective hopping amplitudes $\chi_m$ for each link $m$ between two sites are shown respectively in (e,f,g) and (h) depicts $\Delta_i^p$.} 
    \label{fig:nonint_rho13}
\end{figure}

Symmetry-breaking, absence and presence of topological properties in the interacting and non-interacting $\text{BODW}_{1/3}$ phases are given in Figures~\ref{fig:int_rho13} and~\ref{fig:nonint_rho13} respectively. Figs.~\ref{fig:int_rho13}(a)-(b) depict the site and bond densities and the corresponding structure factors are plotted in Figs.~\ref{fig:int_rho13}(c)-(d). Due to NN and NNN interactions, the interacting $\text{BODW}_{1/3}$ phase possesses an enlarged unit cell with six sites. This differs from the interacting $\text{BODW}_{1/4}$ phase where only the presence of NN interactions is required. The unit cell structure can be seen from Fig.~\ref{fig:int_rho13}(a) where the site density profile exhibits oscillations with period-6, which differs from the non-interacting $\text{BODW}_{1/3}$ with a unit cell of three sites exhibiting site density oscillations with period-3 given in Fig.~\ref{fig:nonint_rho13}(a). Due to the finite particle occupation in certain nearby sites, dimerized hopping gives rise to finite bond density at certain links. This can be seen from Fig.~\ref{fig:int_rho13}(b) where $\braket{\hat B_i}$ makes a peak at the first and the third link in the unit cell. This differs from the non-interacting $\text{BODW}_{1/3}$ where there is a single dimer inside each unit cell as can be seen from Fig.~\ref{fig:nonint_rho13}(b) with a single peak for every three sites. The symmetry-breaking with a unit cell of six sites translates into the peaks of the structure factors $\mathcal{S}_{DW}(k)$ at $k=\pi/3, 2\pi/3$ and $\mathcal{S}_{BO}(k)$ at $k=\pi/3,2\pi/3, \pi$ as shown in Fig.~\ref{fig:int_rho13}(c,d), which is different than the non-interacting $\text{BODW}_{1/3}$ with a unit cell of three sites as displayed in Fig.~\ref{fig:nonint_rho13}(c,d). 

The fact that the interacting model admits a trivial phase independent of whether $|t_1|>|t_2|$ or $|t_1|<|t_2|$ is shown by plotting the entanglement spectrum, local Berry phase, effective hoppings, and edge state population in Figs.~\ref{fig:int_rho13}(e)-(h) respectively. Topological features are absent in the interacting $\text{BODW}_{1/3}$, therefore, the entanglement spectrum does not consist of groups of degenerate Schmidt coefficients as shown in Fig.~\ref{fig:int_rho13}(e). As mentioned in section \ref{sec:phases}, the system in the non-interacting limit at $\rho=1/3$ with periodic modulation of the hopping amplitudes $[t_1,t_2,t_3]$ as $t_i=t_{i+3}$ hosts the SPT phase as long as the constraints $|t_1|=|t_2|$ and $|t_1|<|t_3|$ are satisfied. This is reflected in the entanglement spectrum with degeneracy for the non-interacting $\text{BODW}_{1/3}$ in Fig.~\ref{fig:nonint_rho13}(e). We also realize that the entanglement spectrum degeneracy is not entirely twofold differing from the non-trivial $\text{BODW}_{1/4}$ phase. It consists of groups of degenerate Schmidt coefficients with multiplicities two and four. The effective hopping amplitudes in the long-range interacting model do not obey the hopping constraints as shown in Fig.~\ref{fig:int_rho13}(g). Fig.~\ref{fig:int_rho13}(f) shows that the local Berry phase in the interacting $\text{BODW}_{1/3}$ does not signal for topological features whereas in the non-interacting case, it is quantized to $\pi$ at every inter-cell link as shown in Fig.~\ref{fig:nonint_rho13}(f). Similar to the trivial $\text{BODW}_{1/4}$, the edge state localization in the interacting $\text{BODW}_{1/3}$ is also absent as can be seen from Fig.~\ref{fig:int_rho13}(h). In contrast to  non-trivial $\text{BODW}_{1/4}$ [Fig.~\ref{fig:spt_rho14}(h)], non-interacting topological $\text{BODW}_{1/3}$ makes two peaks at the value $0.5$ at the end of the lattice as shown in Fig.~\ref{fig:nonint_rho13}(h). This can be attributed to having no interactions in the system. Including NN interactions to the system with hoppings with period-3 at the filling $\rho=1/3$ yields similar results with the non-trivial interacting $\text{BODW}_{1/4}$ at the filling $\rho=1/4$ and non-trivial BO phase at the filling $\rho=1/2$.

\section{Conclusion}
This study was initiated in our recent work \cite{zeybek_quantum_2023} where a Rydberg quantum simulator was utilized to probe the interplay between short- and long-range interactions. Such competing processes were shown to host phases with both density wave and bond orders. In this work, we expanded that by contrasting interacting and non-interacting BODW phases.  We then have shown that long-range interactions induce the emergence of superlattice structures with higher periods despite having a dimerized lattice. This was illustrated by showing that the BODW phases with emergent symmetry-breaking properties can be stabilized without the explicit superlattice structure of the non-interacting counterpart. Beyond NN interactions are shown to realize a fundamentally different BODW phase at the filling $\rho=1/3$, in particular, with a different symmetry-breaking pattern where the unit cell size and structure differs from the non-interacting case. Changing the dimerization pattern in the hopping amplitudes is found to realize trivial symmetry-broken sectors of the BODW phases. Including on-site modulations are used to help stabilize topological $\text{BODW}_{1/4}$ phase. Our work provides insights into the interaction-induced emergent ground state properties of long-range interacting hardcore bosons and motivates investigating higher dimensional lattices with more connectivity.

\begin{acknowledgments}
This work is funded by the Cluster of Excellence
		``CUI: Advanced Imaging of Matter'' of the Deutsche Forschungsgemeinschaft (DFG) - EXC 2056 - Project ID 390715994. This work is funded by the German Federal Ministry of Education and Research within the funding program ``quantum technologies - from basic research to market" under contract 13N16138.
\end{acknowledgments}

\bibliographystyle{apsrev4-2}
\bibliography{REF_BODW} 

\begin{thebibliography}{72}%
\makeatletter
\providecommand \@ifxundefined [1]{%
 \@ifx{#1\undefined}
}%
\providecommand \@ifnum [1]{%
 \ifnum #1\expandafter \@firstoftwo
 \else \expandafter \@secondoftwo
 \fi
}%
\providecommand \@ifx [1]{%
 \ifx #1\expandafter \@firstoftwo
 \else \expandafter \@secondoftwo
 \fi
}%
\providecommand \natexlab [1]{#1}%
\providecommand \enquote  [1]{``#1''}%
\providecommand \bibnamefont  [1]{#1}%
\providecommand \bibfnamefont [1]{#1}%
\providecommand \citenamefont [1]{#1}%
\providecommand \href@noop [0]{\@secondoftwo}%
\providecommand \href [0]{\begingroup \@sanitize@url \@href}%
\providecommand \@href[1]{\@@startlink{#1}\@@href}%
\providecommand \@@href[1]{\endgroup#1\@@endlink}%
\providecommand \@sanitize@url [0]{\catcode `\\12\catcode `\$12\catcode `\&12\catcode `\#12\catcode `\^12\catcode `\_12\catcode `\%12\relax}%
\providecommand \@@startlink[1]{}%
\providecommand \@@endlink[0]{}%
\providecommand \url  [0]{\begingroup\@sanitize@url \@url }%
\providecommand \@url [1]{\endgroup\@href {#1}{\urlprefix }}%
\providecommand \urlprefix  [0]{URL }%
\providecommand \Eprint [0]{\href }%
\providecommand \doibase [0]{https://doi.org/}%
\providecommand \selectlanguage [0]{\@gobble}%
\providecommand \bibinfo  [0]{\@secondoftwo}%
\providecommand \bibfield  [0]{\@secondoftwo}%
\providecommand \translation [1]{[#1]}%
\providecommand \BibitemOpen [0]{}%
\providecommand \bibitemStop [0]{}%
\providecommand \bibitemNoStop [0]{.\EOS\space}%
\providecommand \EOS [0]{\spacefactor3000\relax}%
\providecommand \BibitemShut  [1]{\csname bibitem#1\endcsname}%
\let\auto@bib@innerbib\@empty
%</preamble>
\bibitem [{\citenamefont {Giamarchi}(2003)}]{giamarchi_quantum_2003}%
  \BibitemOpen
  \bibfield  {author} {\bibinfo {author} {\bibfnamefont {T.}~\bibnamefont {Giamarchi}},\ }\href {https://doi.org/10.1093/acprof:oso/9780198525004.001.0001} {\emph {\bibinfo {title} {Quantum {Physics} in {One} {Dimension}}}}\ (\bibinfo  {publisher} {Clarendon Press},\ \bibinfo {year} {2003})\BibitemShut {NoStop}%
\bibitem [{\citenamefont {Jaksch}\ \emph {et~al.}(1998)\citenamefont {Jaksch}, \citenamefont {Bruder}, \citenamefont {Cirac}, \citenamefont {Gardiner},\ and\ \citenamefont {Zoller}}]{jaksch98}%
  \BibitemOpen
  \bibfield  {author} {\bibinfo {author} {\bibfnamefont {D.}~\bibnamefont {Jaksch}}, \bibinfo {author} {\bibfnamefont {C.}~\bibnamefont {Bruder}}, \bibinfo {author} {\bibfnamefont {J.~I.}\ \bibnamefont {Cirac}}, \bibinfo {author} {\bibfnamefont {C.~W.}\ \bibnamefont {Gardiner}},\ and\ \bibinfo {author} {\bibfnamefont {P.}~\bibnamefont {Zoller}},\ }\href {https://doi.org/10.1103/PhysRevLett.81.3108} {\bibfield  {journal} {\bibinfo  {journal} {Phys. Rev. Lett.}\ }\textbf {\bibinfo {volume} {81}},\ \bibinfo {pages} {3108} (\bibinfo {year} {1998})}\BibitemShut {NoStop}%
\bibitem [{\citenamefont {Arovas}\ \emph {et~al.}(2022)\citenamefont {Arovas}, \citenamefont {Berg}, \citenamefont {Kivelson},\ and\ \citenamefont {Raghu}}]{arovas}%
  \BibitemOpen
  \bibfield  {author} {\bibinfo {author} {\bibfnamefont {D.~P.}\ \bibnamefont {Arovas}}, \bibinfo {author} {\bibfnamefont {E.}~\bibnamefont {Berg}}, \bibinfo {author} {\bibfnamefont {S.~A.}\ \bibnamefont {Kivelson}},\ and\ \bibinfo {author} {\bibfnamefont {S.}~\bibnamefont {Raghu}},\ }\href {https://doi.org/10.1146/annurev-conmatphys-031620-102024} {\bibfield  {journal} {\bibinfo  {journal} {Annu. Rev. Condens. Matter Phys.}\ }\textbf {\bibinfo {volume} {13}},\ \bibinfo {pages} {239} (\bibinfo {year} {2022})}\BibitemShut {NoStop}%
\bibitem [{\citenamefont {Cazalilla}\ \emph {et~al.}(2011)\citenamefont {Cazalilla}, \citenamefont {Citro}, \citenamefont {Giamarchi}, \citenamefont {Orignac},\ and\ \citenamefont {Rigol}}]{cazalilla_one_2011}%
  \BibitemOpen
  \bibfield  {author} {\bibinfo {author} {\bibfnamefont {M.~A.}\ \bibnamefont {Cazalilla}}, \bibinfo {author} {\bibfnamefont {R.}~\bibnamefont {Citro}}, \bibinfo {author} {\bibfnamefont {T.}~\bibnamefont {Giamarchi}}, \bibinfo {author} {\bibfnamefont {E.}~\bibnamefont {Orignac}},\ and\ \bibinfo {author} {\bibfnamefont {M.}~\bibnamefont {Rigol}},\ }\href {https://doi.org/10.1103/RevModPhys.83.1405} {\bibfield  {journal} {\bibinfo  {journal} {Rev. Mod. Phys.}\ }\textbf {\bibinfo {volume} {83}},\ \bibinfo {pages} {1405} (\bibinfo {year} {2011})}\BibitemShut {NoStop}%
\bibitem [{\citenamefont {Freericks}\ and\ \citenamefont {Monien}(1994)}]{freericks_phase_1994}%
  \BibitemOpen
  \bibfield  {author} {\bibinfo {author} {\bibfnamefont {J.~K.}\ \bibnamefont {Freericks}}\ and\ \bibinfo {author} {\bibfnamefont {H.}~\bibnamefont {Monien}},\ }\href {https://doi.org/10.1209/0295-5075/26/7/012} {\bibfield  {journal} {\bibinfo  {journal} {Europhys. Lett.}\ }\textbf {\bibinfo {volume} {26}},\ \bibinfo {pages} {545} (\bibinfo {year} {1994})}\BibitemShut {NoStop}%
\bibitem [{\citenamefont {Ejima}\ \emph {et~al.}(2011)\citenamefont {Ejima}, \citenamefont {Fehske},\ and\ \citenamefont {Gebhard}}]{ejima_dynamic_2011}%
  \BibitemOpen
  \bibfield  {author} {\bibinfo {author} {\bibfnamefont {S.}~\bibnamefont {Ejima}}, \bibinfo {author} {\bibfnamefont {H.}~\bibnamefont {Fehske}},\ and\ \bibinfo {author} {\bibfnamefont {F.}~\bibnamefont {Gebhard}},\ }\href {https://doi.org/10.1209/0295-5075/93/30002} {\bibfield  {journal} {\bibinfo  {journal} {Europhys. Lett.}\ }\textbf {\bibinfo {volume} {93}},\ \bibinfo {pages} {30002} (\bibinfo {year} {2011})}\BibitemShut {NoStop}%
\bibitem [{\citenamefont {Anders}\ \emph {et~al.}(2010)\citenamefont {Anders}, \citenamefont {Gull}, \citenamefont {Pollet}, \citenamefont {Troyer},\ and\ \citenamefont {Werner}}]{anders2010}%
  \BibitemOpen
  \bibfield  {author} {\bibinfo {author} {\bibfnamefont {P.}~\bibnamefont {Anders}}, \bibinfo {author} {\bibfnamefont {E.}~\bibnamefont {Gull}}, \bibinfo {author} {\bibfnamefont {L.}~\bibnamefont {Pollet}}, \bibinfo {author} {\bibfnamefont {M.}~\bibnamefont {Troyer}},\ and\ \bibinfo {author} {\bibfnamefont {P.}~\bibnamefont {Werner}},\ }\href {https://doi.org/10.1103/PhysRevLett.105.096402} {\bibfield  {journal} {\bibinfo  {journal} {Phys. Rev. Lett.}\ }\textbf {\bibinfo {volume} {105}},\ \bibinfo {pages} {096402} (\bibinfo {year} {2010})}\BibitemShut {NoStop}%
\bibitem [{\citenamefont {Elstner}\ and\ \citenamefont {Monien}(1999)}]{elstner1999}%
  \BibitemOpen
  \bibfield  {author} {\bibinfo {author} {\bibfnamefont {N.}~\bibnamefont {Elstner}}\ and\ \bibinfo {author} {\bibfnamefont {H.}~\bibnamefont {Monien}},\ }\href {https://doi.org/10.1103/PhysRevB.59.12184} {\bibfield  {journal} {\bibinfo  {journal} {Phys. Rev. B}\ }\textbf {\bibinfo {volume} {59}},\ \bibinfo {pages} {12184} (\bibinfo {year} {1999})}\BibitemShut {NoStop}%
\bibitem [{\citenamefont {Essler}\ \emph {et~al.}(2005)\citenamefont {Essler}, \citenamefont {Frahm}, \citenamefont {Göhmann}, \citenamefont {Klümper},\ and\ \citenamefont {Korepin}}]{essler_one-dimensional_2005}%
  \BibitemOpen
  \bibfield  {author} {\bibinfo {author} {\bibfnamefont {F.~H.~L.}\ \bibnamefont {Essler}}, \bibinfo {author} {\bibfnamefont {H.}~\bibnamefont {Frahm}}, \bibinfo {author} {\bibfnamefont {F.}~\bibnamefont {Göhmann}}, \bibinfo {author} {\bibfnamefont {A.}~\bibnamefont {Klümper}},\ and\ \bibinfo {author} {\bibfnamefont {V.~E.}\ \bibnamefont {Korepin}},\ }\href {https://doi.org/10.1017/CBO9780511534843} {\emph {\bibinfo {title} {The One-Dimensional Hubbard Model}}}\ (\bibinfo  {publisher} {Cambridge University Press},\ \bibinfo {year} {2005})\BibitemShut {NoStop}%
\bibitem [{\citenamefont {Batrouni}\ \emph {et~al.}(1995)\citenamefont {Batrouni}, \citenamefont {Scalettar}, \citenamefont {Zimanyi},\ and\ \citenamefont {Kampf}}]{batrouni_supersolids_1995}%
  \BibitemOpen
  \bibfield  {author} {\bibinfo {author} {\bibfnamefont {G.~G.}\ \bibnamefont {Batrouni}}, \bibinfo {author} {\bibfnamefont {R.~T.}\ \bibnamefont {Scalettar}}, \bibinfo {author} {\bibfnamefont {G.~T.}\ \bibnamefont {Zimanyi}},\ and\ \bibinfo {author} {\bibfnamefont {A.~P.}\ \bibnamefont {Kampf}},\ }\href {https://doi.org/10.1103/PhysRevLett.74.2527} {\bibfield  {journal} {\bibinfo  {journal} {Phys. Rev. Lett.}\ }\textbf {\bibinfo {volume} {74}},\ \bibinfo {pages} {2527} (\bibinfo {year} {1995})}\BibitemShut {NoStop}%
\bibitem [{\citenamefont {Bloch}(2005)}]{bloch_ultracold_2005}%
  \BibitemOpen
  \bibfield  {author} {\bibinfo {author} {\bibfnamefont {I.}~\bibnamefont {Bloch}},\ }\href {https://doi.org/10.1038/nphys138} {\bibfield  {journal} {\bibinfo  {journal} {Nat. Phys.}\ }\textbf {\bibinfo {volume} {1}},\ \bibinfo {pages} {23} (\bibinfo {year} {2005})}\BibitemShut {NoStop}%
\bibitem [{\citenamefont {Bloch}\ \emph {et~al.}(2012)\citenamefont {Bloch}, \citenamefont {Dalibard},\ and\ \citenamefont {Nascimb\`ene}}]{bloch_quantum_2012}%
  \BibitemOpen
  \bibfield  {author} {\bibinfo {author} {\bibfnamefont {I.}~\bibnamefont {Bloch}}, \bibinfo {author} {\bibfnamefont {J.}~\bibnamefont {Dalibard}},\ and\ \bibinfo {author} {\bibfnamefont {S.}~\bibnamefont {Nascimb\`ene}},\ }\href {https://doi.org/10.1038/nphys2259} {\bibfield  {journal} {\bibinfo  {journal} {Nat. Phys.}\ }\textbf {\bibinfo {volume} {8}},\ \bibinfo {pages} {267} (\bibinfo {year} {2012})}\BibitemShut {NoStop}%
\bibitem [{\citenamefont {Lewenstein}\ \emph {et~al.}(2007)\citenamefont {Lewenstein}, \citenamefont {Sanpera}, \citenamefont {Ahufinger}, \citenamefont {Damski}, \citenamefont {Sen(De)},\ and\ \citenamefont {Sen}}]{lewens2007}%
  \BibitemOpen
  \bibfield  {author} {\bibinfo {author} {\bibfnamefont {M.}~\bibnamefont {Lewenstein}}, \bibinfo {author} {\bibfnamefont {A.}~\bibnamefont {Sanpera}}, \bibinfo {author} {\bibfnamefont {V.}~\bibnamefont {Ahufinger}}, \bibinfo {author} {\bibfnamefont {B.}~\bibnamefont {Damski}}, \bibinfo {author} {\bibfnamefont {A.}~\bibnamefont {Sen(De)}},\ and\ \bibinfo {author} {\bibfnamefont {U.}~\bibnamefont {Sen}},\ }\href {https://doi.org/10.1080/00018730701223200} {\bibfield  {journal} {\bibinfo  {journal} {Adv. Phys.}\ }\textbf {\bibinfo {volume} {56}},\ \bibinfo {pages} {243} (\bibinfo {year} {2007})}\BibitemShut {NoStop}%
\bibitem [{\citenamefont {Weimer}\ \emph {et~al.}(2010)\citenamefont {Weimer}, \citenamefont {M\"uller}, \citenamefont {Lesanovsky}, \citenamefont {Zoller},\ and\ \citenamefont {B\"uchler}}]{weimer_rydberg_2010}%
  \BibitemOpen
  \bibfield  {author} {\bibinfo {author} {\bibfnamefont {H.}~\bibnamefont {Weimer}}, \bibinfo {author} {\bibfnamefont {M.}~\bibnamefont {M\"uller}}, \bibinfo {author} {\bibfnamefont {I.}~\bibnamefont {Lesanovsky}}, \bibinfo {author} {\bibfnamefont {P.}~\bibnamefont {Zoller}},\ and\ \bibinfo {author} {\bibfnamefont {H.~P.}\ \bibnamefont {B\"uchler}},\ }\href {https://doi.org/10.1038/nphys1614} {\bibfield  {journal} {\bibinfo  {journal} {Nat. Phys.}\ }\textbf {\bibinfo {volume} {6}},\ \bibinfo {pages} {382} (\bibinfo {year} {2010})}\BibitemShut {NoStop}%
\bibitem [{\citenamefont {Browaeys}\ and\ \citenamefont {Lahaye}(2020)}]{browaeys_many-body_2020}%
  \BibitemOpen
  \bibfield  {author} {\bibinfo {author} {\bibfnamefont {A.}~\bibnamefont {Browaeys}}\ and\ \bibinfo {author} {\bibfnamefont {T.}~\bibnamefont {Lahaye}},\ }\href {https://doi.org/10.1038/s41567-019-0733-z} {\bibfield  {journal} {\bibinfo  {journal} {Nat. Phys.}\ }\textbf {\bibinfo {volume} {16}},\ \bibinfo {pages} {132} (\bibinfo {year} {2020})}\BibitemShut {NoStop}%
\bibitem [{\citenamefont {Greiner}\ \emph {et~al.}(2002)\citenamefont {Greiner}, \citenamefont {Mandel}, \citenamefont {Esslinger}, \citenamefont {Hänsch},\ and\ \citenamefont {Bloch}}]{greiner_quantum_2002}%
  \BibitemOpen
  \bibfield  {author} {\bibinfo {author} {\bibfnamefont {M.}~\bibnamefont {Greiner}}, \bibinfo {author} {\bibfnamefont {O.}~\bibnamefont {Mandel}}, \bibinfo {author} {\bibfnamefont {T.}~\bibnamefont {Esslinger}}, \bibinfo {author} {\bibfnamefont {T.~W.}\ \bibnamefont {Hänsch}},\ and\ \bibinfo {author} {\bibfnamefont {I.}~\bibnamefont {Bloch}},\ }\href {https://doi.org/10.1038/415039a} {\bibfield  {journal} {\bibinfo  {journal} {Nature (London)}\ }\textbf {\bibinfo {volume} {415}},\ \bibinfo {pages} {39} (\bibinfo {year} {2002})}\BibitemShut {NoStop}%
\bibitem [{\citenamefont {Esslinger}(2010)}]{esslin2010}%
  \BibitemOpen
  \bibfield  {author} {\bibinfo {author} {\bibfnamefont {T.}~\bibnamefont {Esslinger}},\ }\href {https://doi.org/10.1146/annurev-conmatphys-070909-104059} {\bibfield  {journal} {\bibinfo  {journal} {Annu. Rev. Condens. Matter Phys.}\ }\textbf {\bibinfo {volume} {1}},\ \bibinfo {pages} {129} (\bibinfo {year} {2010})}\BibitemShut {NoStop}%
\bibitem [{\citenamefont {Huber}\ \emph {et~al.}(2007)\citenamefont {Huber}, \citenamefont {Altman}, \citenamefont {B\"uchler},\ and\ \citenamefont {Blatter}}]{huber2007}%
  \BibitemOpen
  \bibfield  {author} {\bibinfo {author} {\bibfnamefont {S.~D.}\ \bibnamefont {Huber}}, \bibinfo {author} {\bibfnamefont {E.}~\bibnamefont {Altman}}, \bibinfo {author} {\bibfnamefont {H.~P.}\ \bibnamefont {B\"uchler}},\ and\ \bibinfo {author} {\bibfnamefont {G.}~\bibnamefont {Blatter}},\ }\href {https://doi.org/10.1103/PhysRevB.75.085106} {\bibfield  {journal} {\bibinfo  {journal} {Phys. Rev. B}\ }\textbf {\bibinfo {volume} {75}},\ \bibinfo {pages} {085106} (\bibinfo {year} {2007})}\BibitemShut {NoStop}%
\bibitem [{\citenamefont {Zwerger}(2003)}]{Wilhelm_2003}%
  \BibitemOpen
  \bibfield  {author} {\bibinfo {author} {\bibfnamefont {W.}~\bibnamefont {Zwerger}},\ }\href {https://doi.org/10.1088/1464-4266/5/2/352} {\bibfield  {journal} {\bibinfo  {journal} {J. Opt. B: Quantum Semiclass. Opt.}\ }\textbf {\bibinfo {volume} {5}},\ \bibinfo {pages} {S9} (\bibinfo {year} {2003})}\BibitemShut {NoStop}%
\bibitem [{\citenamefont {Zhu}\ \emph {et~al.}(2013)\citenamefont {Zhu}, \citenamefont {Wang}, \citenamefont {Chan},\ and\ \citenamefont {Duan}}]{zhu_topological_2013}%
  \BibitemOpen
  \bibfield  {author} {\bibinfo {author} {\bibfnamefont {S.-L.}\ \bibnamefont {Zhu}}, \bibinfo {author} {\bibfnamefont {Z.-D.}\ \bibnamefont {Wang}}, \bibinfo {author} {\bibfnamefont {Y.-H.}\ \bibnamefont {Chan}},\ and\ \bibinfo {author} {\bibfnamefont {L.-M.}\ \bibnamefont {Duan}},\ }\href {https://doi.org/10.1103/PhysRevLett.110.075303} {\bibfield  {journal} {\bibinfo  {journal} {Phys. Rev. Lett.}\ }\textbf {\bibinfo {volume} {110}},\ \bibinfo {pages} {075303} (\bibinfo {year} {2013})}\BibitemShut {NoStop}%
\bibitem [{\citenamefont {Maschler}\ and\ \citenamefont {Ritsch}(2005)}]{maschler2005}%
  \BibitemOpen
  \bibfield  {author} {\bibinfo {author} {\bibfnamefont {C.}~\bibnamefont {Maschler}}\ and\ \bibinfo {author} {\bibfnamefont {H.}~\bibnamefont {Ritsch}},\ }\href {https://doi.org/10.1103/PhysRevLett.95.260401} {\bibfield  {journal} {\bibinfo  {journal} {Phys. Rev. Lett.}\ }\textbf {\bibinfo {volume} {95}},\ \bibinfo {pages} {260401} (\bibinfo {year} {2005})}\BibitemShut {NoStop}%
\bibitem [{\citenamefont {Baier}\ \emph {et~al.}(2016)\citenamefont {Baier}, \citenamefont {Mark}, \citenamefont {Petter}, \citenamefont {Aikawa}, \citenamefont {Chomaz}, \citenamefont {Cai}, \citenamefont {Baranov}, \citenamefont {Zoller},\ and\ \citenamefont {Ferlaino}}]{baier_extended_2016}%
  \BibitemOpen
  \bibfield  {author} {\bibinfo {author} {\bibfnamefont {S.}~\bibnamefont {Baier}}, \bibinfo {author} {\bibfnamefont {M.~J.}\ \bibnamefont {Mark}}, \bibinfo {author} {\bibfnamefont {D.}~\bibnamefont {Petter}}, \bibinfo {author} {\bibfnamefont {K.}~\bibnamefont {Aikawa}}, \bibinfo {author} {\bibfnamefont {L.}~\bibnamefont {Chomaz}}, \bibinfo {author} {\bibfnamefont {Z.}~\bibnamefont {Cai}}, \bibinfo {author} {\bibfnamefont {M.}~\bibnamefont {Baranov}}, \bibinfo {author} {\bibfnamefont {P.}~\bibnamefont {Zoller}},\ and\ \bibinfo {author} {\bibfnamefont {F.}~\bibnamefont {Ferlaino}},\ }\href {https://doi.org/10.1126/science.aac9812} {\bibfield  {journal} {\bibinfo  {journal} {Science}\ }\textbf {\bibinfo {volume} {352}},\ \bibinfo {pages} {201} (\bibinfo {year} {2016})}\BibitemShut {NoStop}%
\bibitem [{\citenamefont {Trefzger}\ \emph {et~al.}(2011)\citenamefont {Trefzger}, \citenamefont {Menotti}, \citenamefont {Capogrosso-Sansone},\ and\ \citenamefont {Lewenstein}}]{Trefzger_2011}%
  \BibitemOpen
  \bibfield  {author} {\bibinfo {author} {\bibfnamefont {C.}~\bibnamefont {Trefzger}}, \bibinfo {author} {\bibfnamefont {C.}~\bibnamefont {Menotti}}, \bibinfo {author} {\bibfnamefont {B.}~\bibnamefont {Capogrosso-Sansone}},\ and\ \bibinfo {author} {\bibfnamefont {M.}~\bibnamefont {Lewenstein}},\ }\href {https://doi.org/10.1088/0953-4075/44/19/193001} {\bibfield  {journal} {\bibinfo  {journal} {J. Phys. B: At. Mol. Opt. Phys.}\ }\textbf {\bibinfo {volume} {44}},\ \bibinfo {pages} {193001} (\bibinfo {year} {2011})}\BibitemShut {NoStop}%
\bibitem [{\citenamefont {Rossini}\ and\ \citenamefont {Fazio}(2012)}]{rossini_phase_2012}%
  \BibitemOpen
  \bibfield  {author} {\bibinfo {author} {\bibfnamefont {D.}~\bibnamefont {Rossini}}\ and\ \bibinfo {author} {\bibfnamefont {R.}~\bibnamefont {Fazio}},\ }\href {https://doi.org/10.1088/1367-2630/14/6/065012} {\bibfield  {journal} {\bibinfo  {journal} {New J. Phys.}\ }\textbf {\bibinfo {volume} {14}},\ \bibinfo {pages} {065012} (\bibinfo {year} {2012})}\BibitemShut {NoStop}%
\bibitem [{\citenamefont {Mazzarella}\ \emph {et~al.}(2006)\citenamefont {Mazzarella}, \citenamefont {Giampaolo},\ and\ \citenamefont {Illuminati}}]{Mazz2006EBHM}%
  \BibitemOpen
  \bibfield  {author} {\bibinfo {author} {\bibfnamefont {G.}~\bibnamefont {Mazzarella}}, \bibinfo {author} {\bibfnamefont {S.~M.}\ \bibnamefont {Giampaolo}},\ and\ \bibinfo {author} {\bibfnamefont {F.}~\bibnamefont {Illuminati}},\ }\href {https://doi.org/10.1103/PhysRevA.73.013625} {\bibfield  {journal} {\bibinfo  {journal} {Phys. Rev. A}\ }\textbf {\bibinfo {volume} {73}},\ \bibinfo {pages} {013625} (\bibinfo {year} {2006})}\BibitemShut {NoStop}%
\bibitem [{\citenamefont {Kraus}\ \emph {et~al.}(2020)\citenamefont {Kraus}, \citenamefont {Biedroń}, \citenamefont {Zakrzewski},\ and\ \citenamefont {Morigi}}]{kraus_superfluid_2020}%
  \BibitemOpen
  \bibfield  {author} {\bibinfo {author} {\bibfnamefont {R.}~\bibnamefont {Kraus}}, \bibinfo {author} {\bibfnamefont {K.}~\bibnamefont {Biedroń}}, \bibinfo {author} {\bibfnamefont {J.}~\bibnamefont {Zakrzewski}},\ and\ \bibinfo {author} {\bibfnamefont {G.}~\bibnamefont {Morigi}},\ }\href {https://doi.org/10.1103/PhysRevB.101.174505} {\bibfield  {journal} {\bibinfo  {journal} {Phys. Rev. B}\ }\textbf {\bibinfo {volume} {101}},\ \bibinfo {pages} {174505} (\bibinfo {year} {2020})}\BibitemShut {NoStop}%
\bibitem [{\citenamefont {Mishra}\ \emph {et~al.}(2009{\natexlab{a}})\citenamefont {Mishra}, \citenamefont {Pai}, \citenamefont {Ramanan}, \citenamefont {Luthra},\ and\ \citenamefont {Das}}]{mishra_supersolid_2009}%
  \BibitemOpen
  \bibfield  {author} {\bibinfo {author} {\bibfnamefont {T.}~\bibnamefont {Mishra}}, \bibinfo {author} {\bibfnamefont {R.~V.}\ \bibnamefont {Pai}}, \bibinfo {author} {\bibfnamefont {S.}~\bibnamefont {Ramanan}}, \bibinfo {author} {\bibfnamefont {M.~S.}\ \bibnamefont {Luthra}},\ and\ \bibinfo {author} {\bibfnamefont {B.~P.}\ \bibnamefont {Das}},\ }\href {https://doi.org/10.1103/PhysRevA.80.043614} {\bibfield  {journal} {\bibinfo  {journal} {Phys. Rev. A}\ }\textbf {\bibinfo {volume} {80}},\ \bibinfo {pages} {043614} (\bibinfo {year} {2009}{\natexlab{a}})}\BibitemShut {NoStop}%
\bibitem [{\citenamefont {Dalla~Torre}\ \emph {et~al.}(2006)\citenamefont {Dalla~Torre}, \citenamefont {Berg},\ and\ \citenamefont {Altman}}]{dalla_torre_hidden_2006}%
  \BibitemOpen
  \bibfield  {author} {\bibinfo {author} {\bibfnamefont {E.~G.}\ \bibnamefont {Dalla~Torre}}, \bibinfo {author} {\bibfnamefont {E.}~\bibnamefont {Berg}},\ and\ \bibinfo {author} {\bibfnamefont {E.}~\bibnamefont {Altman}},\ }\href {https://doi.org/10.1103/PhysRevLett.97.260401} {\bibfield  {journal} {\bibinfo  {journal} {Phys. Rev. Lett.}\ }\textbf {\bibinfo {volume} {97}},\ \bibinfo {pages} {260401} (\bibinfo {year} {2006})}\BibitemShut {NoStop}%
\bibitem [{\citenamefont {Berg}\ \emph {et~al.}(2008)\citenamefont {Berg}, \citenamefont {Dalla~Torre}, \citenamefont {Giamarchi},\ and\ \citenamefont {Altman}}]{berg_rise_2008}%
  \BibitemOpen
  \bibfield  {author} {\bibinfo {author} {\bibfnamefont {E.}~\bibnamefont {Berg}}, \bibinfo {author} {\bibfnamefont {E.~G.}\ \bibnamefont {Dalla~Torre}}, \bibinfo {author} {\bibfnamefont {T.}~\bibnamefont {Giamarchi}},\ and\ \bibinfo {author} {\bibfnamefont {E.}~\bibnamefont {Altman}},\ }\href {https://doi.org/10.1103/PhysRevB.77.245119} {\bibfield  {journal} {\bibinfo  {journal} {Phys. Rev. B}\ }\textbf {\bibinfo {volume} {77}},\ \bibinfo {pages} {245119} (\bibinfo {year} {2008})}\BibitemShut {NoStop}%
\bibitem [{\citenamefont {Kottmann}\ \emph {et~al.}(2021)\citenamefont {Kottmann}, \citenamefont {Haller}, \citenamefont {Ac\'{\i}n}, \citenamefont {Astrakharchik},\ and\ \citenamefont {Lewenstein}}]{kottmann_supersolid-superfluid_2021}%
  \BibitemOpen
  \bibfield  {author} {\bibinfo {author} {\bibfnamefont {K.}~\bibnamefont {Kottmann}}, \bibinfo {author} {\bibfnamefont {A.}~\bibnamefont {Haller}}, \bibinfo {author} {\bibfnamefont {A.}~\bibnamefont {Ac\'{\i}n}}, \bibinfo {author} {\bibfnamefont {G.~E.}\ \bibnamefont {Astrakharchik}},\ and\ \bibinfo {author} {\bibfnamefont {M.}~\bibnamefont {Lewenstein}},\ }\href {https://doi.org/10.1103/PhysRevB.104.174514} {\bibfield  {journal} {\bibinfo  {journal} {Phys. Rev. B}\ }\textbf {\bibinfo {volume} {104}},\ \bibinfo {pages} {174514} (\bibinfo {year} {2021})}\BibitemShut {NoStop}%
\bibitem [{\citenamefont {Roth}\ and\ \citenamefont {Burnett}(2003)}]{roth_phase_2003}%
  \BibitemOpen
  \bibfield  {author} {\bibinfo {author} {\bibfnamefont {R.}~\bibnamefont {Roth}}\ and\ \bibinfo {author} {\bibfnamefont {K.}~\bibnamefont {Burnett}},\ }\href {https://doi.org/10.1103/PhysRevA.68.023604} {\bibfield  {journal} {\bibinfo  {journal} {Phys. Rev. A}\ }\textbf {\bibinfo {volume} {68}},\ \bibinfo {pages} {023604} (\bibinfo {year} {2003})}\BibitemShut {NoStop}%
\bibitem [{\citenamefont {Sugimoto}\ \emph {et~al.}(2019)\citenamefont {Sugimoto}, \citenamefont {Ejima}, \citenamefont {Lange},\ and\ \citenamefont {Fehske}}]{sugimoto_quantum_2019}%
  \BibitemOpen
  \bibfield  {author} {\bibinfo {author} {\bibfnamefont {K.}~\bibnamefont {Sugimoto}}, \bibinfo {author} {\bibfnamefont {S.}~\bibnamefont {Ejima}}, \bibinfo {author} {\bibfnamefont {F.}~\bibnamefont {Lange}},\ and\ \bibinfo {author} {\bibfnamefont {H.}~\bibnamefont {Fehske}},\ }\href {https://doi.org/10.1103/PhysRevA.99.012122} {\bibfield  {journal} {\bibinfo  {journal} {Phys. Rev. A}\ }\textbf {\bibinfo {volume} {99}},\ \bibinfo {pages} {012122} (\bibinfo {year} {2019})}\BibitemShut {NoStop}%
\bibitem [{\citenamefont {Fraxanet}\ \emph {et~al.}(2022)\citenamefont {Fraxanet}, \citenamefont {González-Cuadra}, \citenamefont {Pfau}, \citenamefont {Lewenstein}, \citenamefont {Langen},\ and\ \citenamefont {Barbiero}}]{fraxanet_topological_2022}%
  \BibitemOpen
  \bibfield  {author} {\bibinfo {author} {\bibfnamefont {J.}~\bibnamefont {Fraxanet}}, \bibinfo {author} {\bibfnamefont {D.}~\bibnamefont {González-Cuadra}}, \bibinfo {author} {\bibfnamefont {T.}~\bibnamefont {Pfau}}, \bibinfo {author} {\bibfnamefont {M.}~\bibnamefont {Lewenstein}}, \bibinfo {author} {\bibfnamefont {T.}~\bibnamefont {Langen}},\ and\ \bibinfo {author} {\bibfnamefont {L.}~\bibnamefont {Barbiero}},\ }\href {https://doi.org/10.1103/PhysRevLett.128.043402} {\bibfield  {journal} {\bibinfo  {journal} {Phys. Rev. Lett.}\ }\textbf {\bibinfo {volume} {128}},\ \bibinfo {pages} {043402} (\bibinfo {year} {2022})}\BibitemShut {NoStop}%
\bibitem [{\citenamefont {Hayashi}\ \emph {et~al.}(2022)\citenamefont {Hayashi}, \citenamefont {Mondal}, \citenamefont {Mishra},\ and\ \citenamefont {Das}}]{hayashi_competing_2022}%
  \BibitemOpen
  \bibfield  {author} {\bibinfo {author} {\bibfnamefont {A.}~\bibnamefont {Hayashi}}, \bibinfo {author} {\bibfnamefont {S.}~\bibnamefont {Mondal}}, \bibinfo {author} {\bibfnamefont {T.}~\bibnamefont {Mishra}},\ and\ \bibinfo {author} {\bibfnamefont {B.~P.}\ \bibnamefont {Das}},\ }\href {https://doi.org/10.1103/PhysRevA.106.013313} {\bibfield  {journal} {\bibinfo  {journal} {Phys. Rev. A}\ }\textbf {\bibinfo {volume} {106}},\ \bibinfo {pages} {013313} (\bibinfo {year} {2022})}\BibitemShut {NoStop}%
\bibitem [{\citenamefont {Zeybek}\ \emph {et~al.}(2023)\citenamefont {Zeybek}, \citenamefont {Mukherjee},\ and\ \citenamefont {Schmelcher}}]{zeybek_quantum_2023}%
  \BibitemOpen
  \bibfield  {author} {\bibinfo {author} {\bibfnamefont {Z.}~\bibnamefont {Zeybek}}, \bibinfo {author} {\bibfnamefont {R.}~\bibnamefont {Mukherjee}},\ and\ \bibinfo {author} {\bibfnamefont {P.}~\bibnamefont {Schmelcher}},\ }\href {https://doi.org/10.1103/PhysRevLett.131.203003} {\bibfield  {journal} {\bibinfo  {journal} {Phys. Rev. Lett.}\ }\textbf {\bibinfo {volume} {131}},\ \bibinfo {pages} {203003} (\bibinfo {year} {2023})}\BibitemShut {NoStop}%
\bibitem [{\citenamefont {Guo}\ and\ \citenamefont {Chen}(2015)}]{guo_kaleidoscope_2015}%
  \BibitemOpen
  \bibfield  {author} {\bibinfo {author} {\bibfnamefont {H.}~\bibnamefont {Guo}}\ and\ \bibinfo {author} {\bibfnamefont {S.}~\bibnamefont {Chen}},\ }\href {https://doi.org/10.1103/PhysRevB.91.041402} {\bibfield  {journal} {\bibinfo  {journal} {Phys. Rev. B}\ }\textbf {\bibinfo {volume} {91}},\ \bibinfo {pages} {041402} (\bibinfo {year} {2015})}\BibitemShut {NoStop}%
\bibitem [{\citenamefont {Senthil}(2015)}]{noauthor_symmetry-protected_nodate}%
  \BibitemOpen
  \bibfield  {author} {\bibinfo {author} {\bibfnamefont {T.}~\bibnamefont {Senthil}},\ }\href {https://doi.org/10.1146/annurev-conmatphys-031214-014740} {\bibfield  {journal} {\bibinfo  {journal} {Annu. Rev. Condens. Matter Phys.}\ }\textbf {\bibinfo {volume} {6}},\ \bibinfo {pages} {299} (\bibinfo {year} {2015})}\BibitemShut {NoStop}%
\bibitem [{\citenamefont {Julià-Farré}\ \emph {et~al.}(2022)\citenamefont {Julià-Farré}, \citenamefont {González-Cuadra}, \citenamefont {Patscheider}, \citenamefont {Mark}, \citenamefont {Ferlaino}, \citenamefont {Lewenstein}, \citenamefont {Barbiero},\ and\ \citenamefont {Dauphin}}]{julia-farre_revealing_2022}%
  \BibitemOpen
  \bibfield  {author} {\bibinfo {author} {\bibfnamefont {S.}~\bibnamefont {Julià-Farré}}, \bibinfo {author} {\bibfnamefont {D.}~\bibnamefont {González-Cuadra}}, \bibinfo {author} {\bibfnamefont {A.}~\bibnamefont {Patscheider}}, \bibinfo {author} {\bibfnamefont {M.~J.}\ \bibnamefont {Mark}}, \bibinfo {author} {\bibfnamefont {F.}~\bibnamefont {Ferlaino}}, \bibinfo {author} {\bibfnamefont {M.}~\bibnamefont {Lewenstein}}, \bibinfo {author} {\bibfnamefont {L.}~\bibnamefont {Barbiero}},\ and\ \bibinfo {author} {\bibfnamefont {A.}~\bibnamefont {Dauphin}},\ }\href {https://doi.org/10.1103/PhysRevResearch.4.L032005} {\bibfield  {journal} {\bibinfo  {journal} {Phys. Rev. Res.}\ }\textbf {\bibinfo {volume} {4}},\ \bibinfo {pages} {L032005} (\bibinfo {year} {2022})}\BibitemShut {NoStop}%
\bibitem [{\citenamefont {Gonz\'alez-Cuadra}\ \emph {et~al.}(2019)\citenamefont {Gonz\'alez-Cuadra}, \citenamefont {Dauphin}, \citenamefont {Grzybowski}, \citenamefont {W\'ojcik}, \citenamefont {Lewenstein},\ and\ \citenamefont {Bermudez}}]{gonzalez-cuadra_symmetry-breaking_2019}%
  \BibitemOpen
  \bibfield  {author} {\bibinfo {author} {\bibfnamefont {D.}~\bibnamefont {Gonz\'alez-Cuadra}}, \bibinfo {author} {\bibfnamefont {A.}~\bibnamefont {Dauphin}}, \bibinfo {author} {\bibfnamefont {P.~R.}\ \bibnamefont {Grzybowski}}, \bibinfo {author} {\bibfnamefont {P.}~\bibnamefont {W\'ojcik}}, \bibinfo {author} {\bibfnamefont {M.}~\bibnamefont {Lewenstein}},\ and\ \bibinfo {author} {\bibfnamefont {A.}~\bibnamefont {Bermudez}},\ }\href {https://doi.org/10.1103/PhysRevB.99.045139} {\bibfield  {journal} {\bibinfo  {journal} {Phys. Rev. B}\ }\textbf {\bibinfo {volume} {99}},\ \bibinfo {pages} {045139} (\bibinfo {year} {2019})}\BibitemShut {NoStop}%
\bibitem [{\citenamefont {González-Cuadra}\ \emph {et~al.}(2019)\citenamefont {González-Cuadra}, \citenamefont {Bermudez}, \citenamefont {Grzybowski}, \citenamefont {Lewenstein},\ and\ \citenamefont {Dauphin}}]{gonzalez-cuadra_intertwined_2019}%
  \BibitemOpen
  \bibfield  {author} {\bibinfo {author} {\bibfnamefont {D.}~\bibnamefont {González-Cuadra}}, \bibinfo {author} {\bibfnamefont {A.}~\bibnamefont {Bermudez}}, \bibinfo {author} {\bibfnamefont {P.~R.}\ \bibnamefont {Grzybowski}}, \bibinfo {author} {\bibfnamefont {M.}~\bibnamefont {Lewenstein}},\ and\ \bibinfo {author} {\bibfnamefont {A.}~\bibnamefont {Dauphin}},\ }\href {https://doi.org/10.1038/s41467-019-10796-8} {\bibfield  {journal} {\bibinfo  {journal} {Nat. Comm.}\ }\textbf {\bibinfo {volume} {10}},\ \bibinfo {pages} {2694} (\bibinfo {year} {2019})}\BibitemShut {NoStop}%
\bibitem [{\citenamefont {Klitzing}\ \emph {et~al.}(1980)\citenamefont {Klitzing}, \citenamefont {Dorda},\ and\ \citenamefont {Pepper}}]{qhall}%
  \BibitemOpen
  \bibfield  {author} {\bibinfo {author} {\bibfnamefont {K.~v.}\ \bibnamefont {Klitzing}}, \bibinfo {author} {\bibfnamefont {G.}~\bibnamefont {Dorda}},\ and\ \bibinfo {author} {\bibfnamefont {M.}~\bibnamefont {Pepper}},\ }\href {https://doi.org/10.1103/PhysRevLett.45.494} {\bibfield  {journal} {\bibinfo  {journal} {Phys. Rev. Lett.}\ }\textbf {\bibinfo {volume} {45}},\ \bibinfo {pages} {494} (\bibinfo {year} {1980})}\BibitemShut {NoStop}%
\bibitem [{\citenamefont {Thouless}\ \emph {et~al.}(1982)\citenamefont {Thouless}, \citenamefont {Kohmoto}, \citenamefont {Nightingale},\ and\ \citenamefont {den Nijs}}]{quantconduc}%
  \BibitemOpen
  \bibfield  {author} {\bibinfo {author} {\bibfnamefont {D.~J.}\ \bibnamefont {Thouless}}, \bibinfo {author} {\bibfnamefont {M.}~\bibnamefont {Kohmoto}}, \bibinfo {author} {\bibfnamefont {M.~P.}\ \bibnamefont {Nightingale}},\ and\ \bibinfo {author} {\bibfnamefont {M.}~\bibnamefont {den Nijs}},\ }\href {https://doi.org/10.1103/PhysRevLett.49.405} {\bibfield  {journal} {\bibinfo  {journal} {Phys. Rev. Lett.}\ }\textbf {\bibinfo {volume} {49}},\ \bibinfo {pages} {405} (\bibinfo {year} {1982})}\BibitemShut {NoStop}%
\bibitem [{\citenamefont {Chiu}\ \emph {et~al.}(2016)\citenamefont {Chiu}, \citenamefont {Teo}, \citenamefont {Schnyder},\ and\ \citenamefont {Ryu}}]{chiu_classification_2016}%
  \BibitemOpen
  \bibfield  {author} {\bibinfo {author} {\bibfnamefont {C.-K.}\ \bibnamefont {Chiu}}, \bibinfo {author} {\bibfnamefont {J.~C.~Y.}\ \bibnamefont {Teo}}, \bibinfo {author} {\bibfnamefont {A.~P.}\ \bibnamefont {Schnyder}},\ and\ \bibinfo {author} {\bibfnamefont {S.}~\bibnamefont {Ryu}},\ }\href {https://doi.org/10.1103/RevModPhys.88.035005} {\bibfield  {journal} {\bibinfo  {journal} {Rev. Mod. Phys.}\ }\textbf {\bibinfo {volume} {88}},\ \bibinfo {pages} {035005} (\bibinfo {year} {2016})}\BibitemShut {NoStop}%
\bibitem [{\citenamefont {Hasan}\ and\ \citenamefont {Kane}(2010)}]{hasan2010}%
  \BibitemOpen
  \bibfield  {author} {\bibinfo {author} {\bibfnamefont {M.~Z.}\ \bibnamefont {Hasan}}\ and\ \bibinfo {author} {\bibfnamefont {C.~L.}\ \bibnamefont {Kane}},\ }\href {https://doi.org/10.1103/RevModPhys.82.3045} {\bibfield  {journal} {\bibinfo  {journal} {Rev. Mod. Phys.}\ }\textbf {\bibinfo {volume} {82}},\ \bibinfo {pages} {3045} (\bibinfo {year} {2010})}\BibitemShut {NoStop}%
\bibitem [{\citenamefont {Bansil}\ \emph {et~al.}(2016)\citenamefont {Bansil}, \citenamefont {Lin},\ and\ \citenamefont {Das}}]{bansil_colloquium_2016}%
  \BibitemOpen
  \bibfield  {author} {\bibinfo {author} {\bibfnamefont {A.}~\bibnamefont {Bansil}}, \bibinfo {author} {\bibfnamefont {H.}~\bibnamefont {Lin}},\ and\ \bibinfo {author} {\bibfnamefont {T.}~\bibnamefont {Das}},\ }\href {https://doi.org/10.1103/RevModPhys.88.021004} {\bibfield  {journal} {\bibinfo  {journal} {Rev. Mod. Phys.}\ }\textbf {\bibinfo {volume} {88}},\ \bibinfo {pages} {021004} (\bibinfo {year} {2016})}\BibitemShut {NoStop}%
\bibitem [{\citenamefont {Chen}\ \emph {et~al.}(2010)\citenamefont {Chen}, \citenamefont {Kou}, \citenamefont {Zhang},\ and\ \citenamefont {Chen}}]{chen_quantum_2010}%
  \BibitemOpen
  \bibfield  {author} {\bibinfo {author} {\bibfnamefont {B.-L.}\ \bibnamefont {Chen}}, \bibinfo {author} {\bibfnamefont {S.-P.}\ \bibnamefont {Kou}}, \bibinfo {author} {\bibfnamefont {Y.}~\bibnamefont {Zhang}},\ and\ \bibinfo {author} {\bibfnamefont {S.}~\bibnamefont {Chen}},\ }\href {https://doi.org/10.1103/PhysRevA.81.053608} {\bibfield  {journal} {\bibinfo  {journal} {Phys. Rev. A}\ }\textbf {\bibinfo {volume} {81}},\ \bibinfo {pages} {053608} (\bibinfo {year} {2010})}\BibitemShut {NoStop}%
\bibitem [{\citenamefont {Bergholtz}\ and\ \citenamefont {Liu}(2013)}]{bergholtz_topological_2013}%
  \BibitemOpen
  \bibfield  {author} {\bibinfo {author} {\bibfnamefont {E.~J.}\ \bibnamefont {Bergholtz}}\ and\ \bibinfo {author} {\bibfnamefont {Z.}~\bibnamefont {Liu}},\ }\href {https://doi.org/10.1142/S021797921330017X} {\bibfield  {journal} {\bibinfo  {journal} {Int. J. Mod. Phys. B}\ }\textbf {\bibinfo {volume} {27}},\ \bibinfo {pages} {1330017} (\bibinfo {year} {2013})}\BibitemShut {NoStop}%
\bibitem [{\citenamefont {Salerno}\ \emph {et~al.}(2020)\citenamefont {Salerno}, \citenamefont {Palumbo}, \citenamefont {Goldman},\ and\ \citenamefont {Di~Liberto}}]{salerno_interaction-induced_2020}%
  \BibitemOpen
  \bibfield  {author} {\bibinfo {author} {\bibfnamefont {G.}~\bibnamefont {Salerno}}, \bibinfo {author} {\bibfnamefont {G.}~\bibnamefont {Palumbo}}, \bibinfo {author} {\bibfnamefont {N.}~\bibnamefont {Goldman}},\ and\ \bibinfo {author} {\bibfnamefont {M.}~\bibnamefont {Di~Liberto}},\ }\href {https://doi.org/10.1103/PhysRevResearch.2.013348} {\bibfield  {journal} {\bibinfo  {journal} {Phys. Rev. Res.}\ }\textbf {\bibinfo {volume} {2}},\ \bibinfo {pages} {013348} (\bibinfo {year} {2020})}\BibitemShut {NoStop}%
\bibitem [{\citenamefont {Rachel}(2018)}]{rachel_interacting_2018}%
  \BibitemOpen
  \bibfield  {author} {\bibinfo {author} {\bibfnamefont {S.}~\bibnamefont {Rachel}},\ }\href {https://doi.org/10.1088/1361-6633/aad6a6} {\bibfield  {journal} {\bibinfo  {journal} {Rep. Prog. Phys.}\ }\textbf {\bibinfo {volume} {81}},\ \bibinfo {pages} {116501} (\bibinfo {year} {2018})}\BibitemShut {NoStop}%
\bibitem [{\citenamefont {Marques}\ and\ \citenamefont {Dias}(2017)}]{marques_multihole_2017}%
  \BibitemOpen
  \bibfield  {author} {\bibinfo {author} {\bibfnamefont {A.~M.}\ \bibnamefont {Marques}}\ and\ \bibinfo {author} {\bibfnamefont {R.~G.}\ \bibnamefont {Dias}},\ }\href {https://doi.org/10.1103/PhysRevB.95.115443} {\bibfield  {journal} {\bibinfo  {journal} {Phys. Rev. B}\ }\textbf {\bibinfo {volume} {95}},\ \bibinfo {pages} {115443} (\bibinfo {year} {2017})}\BibitemShut {NoStop}%
\bibitem [{\citenamefont {Grusdt}\ \emph {et~al.}(2013)\citenamefont {Grusdt}, \citenamefont {Höning},\ and\ \citenamefont {Fleischhauer}}]{grusdt_topological_2013}%
  \BibitemOpen
  \bibfield  {author} {\bibinfo {author} {\bibfnamefont {F.}~\bibnamefont {Grusdt}}, \bibinfo {author} {\bibfnamefont {M.}~\bibnamefont {Höning}},\ and\ \bibinfo {author} {\bibfnamefont {M.}~\bibnamefont {Fleischhauer}},\ }\href {https://doi.org/10.1103/PhysRevLett.110.260405} {\bibfield  {journal} {\bibinfo  {journal} {Phys. Rev. Lett.}\ }\textbf {\bibinfo {volume} {110}},\ \bibinfo {pages} {260405} (\bibinfo {year} {2013})}\BibitemShut {NoStop}%
\bibitem [{\citenamefont {Su}\ \emph {et~al.}(1979)\citenamefont {Su}, \citenamefont {Schrieffer},\ and\ \citenamefont {Heeger}}]{su_solitons_1979}%
  \BibitemOpen
  \bibfield  {author} {\bibinfo {author} {\bibfnamefont {W.~P.}\ \bibnamefont {Su}}, \bibinfo {author} {\bibfnamefont {J.~R.}\ \bibnamefont {Schrieffer}},\ and\ \bibinfo {author} {\bibfnamefont {A.~J.}\ \bibnamefont {Heeger}},\ }\href {https://doi.org/10.1103/PhysRevLett.42.1698} {\bibfield  {journal} {\bibinfo  {journal} {Phys. Rev. Lett.}\ }\textbf {\bibinfo {volume} {42}},\ \bibinfo {pages} {1698} (\bibinfo {year} {1979})}\BibitemShut {NoStop}%
\bibitem [{\citenamefont {Rousseau}\ \emph {et~al.}(2006)\citenamefont {Rousseau}, \citenamefont {Arovas}, \citenamefont {Rigol}, \citenamefont {H\'ebert}, \citenamefont {Batrouni},\ and\ \citenamefont {Scalettar}}]{PhysRevB.73.174516}%
  \BibitemOpen
  \bibfield  {author} {\bibinfo {author} {\bibfnamefont {V.~G.}\ \bibnamefont {Rousseau}}, \bibinfo {author} {\bibfnamefont {D.~P.}\ \bibnamefont {Arovas}}, \bibinfo {author} {\bibfnamefont {M.}~\bibnamefont {Rigol}}, \bibinfo {author} {\bibfnamefont {F.}~\bibnamefont {H\'ebert}}, \bibinfo {author} {\bibfnamefont {G.~G.}\ \bibnamefont {Batrouni}},\ and\ \bibinfo {author} {\bibfnamefont {R.~T.}\ \bibnamefont {Scalettar}},\ }\href {https://doi.org/10.1103/PhysRevB.73.174516} {\bibfield  {journal} {\bibinfo  {journal} {Phys. Rev. B}\ }\textbf {\bibinfo {volume} {73}},\ \bibinfo {pages} {174516} (\bibinfo {year} {2006})}\BibitemShut {NoStop}%
\bibitem [{\citenamefont {Pollmann}\ \emph {et~al.}(2010)\citenamefont {Pollmann}, \citenamefont {Turner}, \citenamefont {Berg},\ and\ \citenamefont {Oshikawa}}]{pollmann_entanglement_2010}%
  \BibitemOpen
  \bibfield  {author} {\bibinfo {author} {\bibfnamefont {F.}~\bibnamefont {Pollmann}}, \bibinfo {author} {\bibfnamefont {A.~M.}\ \bibnamefont {Turner}}, \bibinfo {author} {\bibfnamefont {E.}~\bibnamefont {Berg}},\ and\ \bibinfo {author} {\bibfnamefont {M.}~\bibnamefont {Oshikawa}},\ }\href {https://doi.org/10.1103/PhysRevB.81.064439} {\bibfield  {journal} {\bibinfo  {journal} {Phys. Rev. B}\ }\textbf {\bibinfo {volume} {81}},\ \bibinfo {pages} {064439} (\bibinfo {year} {2010})}\BibitemShut {NoStop}%
\bibitem [{\citenamefont {Pollmann}\ and\ \citenamefont {Turner}(2012)}]{pollmann_detection_2012}%
  \BibitemOpen
  \bibfield  {author} {\bibinfo {author} {\bibfnamefont {F.}~\bibnamefont {Pollmann}}\ and\ \bibinfo {author} {\bibfnamefont {A.~M.}\ \bibnamefont {Turner}},\ }\href {https://doi.org/10.1103/PhysRevB.86.125441} {\bibfield  {journal} {\bibinfo  {journal} {Phys. Rev. B}\ }\textbf {\bibinfo {volume} {86}},\ \bibinfo {pages} {125441} (\bibinfo {year} {2012})}\BibitemShut {NoStop}%
\bibitem [{\citenamefont {Hatsugai}(2006)}]{hatsugai_quantized_2006}%
  \BibitemOpen
  \bibfield  {author} {\bibinfo {author} {\bibfnamefont {Y.}~\bibnamefont {Hatsugai}},\ }\href {https://doi.org/10.1143/JPSJ.75.123601} {\bibfield  {journal} {\bibinfo  {journal} {J. Phys. Soc. Jpn.}\ }\textbf {\bibinfo {volume} {75}},\ \bibinfo {pages} {123601} (\bibinfo {year} {2006})}\BibitemShut {NoStop}%
\bibitem [{\citenamefont {Hirano}\ \emph {et~al.}(2008)\citenamefont {Hirano}, \citenamefont {Katsura},\ and\ \citenamefont {Hatsugai}}]{hatsu2}%
  \BibitemOpen
  \bibfield  {author} {\bibinfo {author} {\bibfnamefont {T.}~\bibnamefont {Hirano}}, \bibinfo {author} {\bibfnamefont {H.}~\bibnamefont {Katsura}},\ and\ \bibinfo {author} {\bibfnamefont {Y.}~\bibnamefont {Hatsugai}},\ }\href {https://doi.org/10.1103/PhysRevB.77.094431} {\bibfield  {journal} {\bibinfo  {journal} {Phys. Rev. B}\ }\textbf {\bibinfo {volume} {77}},\ \bibinfo {pages} {094431} (\bibinfo {year} {2008})}\BibitemShut {NoStop}%
\bibitem [{\citenamefont {Fraxanet}\ \emph {et~al.}(2023)\citenamefont {Fraxanet}, \citenamefont {Dauphin}, \citenamefont {Lewenstein}, \citenamefont {Barbiero},\ and\ \citenamefont {Gonz\'alez-Cuadra}}]{Dynamic_BHSSH_Fraxanet_2023}%
  \BibitemOpen
  \bibfield  {author} {\bibinfo {author} {\bibfnamefont {J.}~\bibnamefont {Fraxanet}}, \bibinfo {author} {\bibfnamefont {A.}~\bibnamefont {Dauphin}}, \bibinfo {author} {\bibfnamefont {M.}~\bibnamefont {Lewenstein}}, \bibinfo {author} {\bibfnamefont {L.}~\bibnamefont {Barbiero}},\ and\ \bibinfo {author} {\bibfnamefont {D.}~\bibnamefont {Gonz\'alez-Cuadra}},\ }\href {https://doi.org/10.1103/PhysRevLett.131.263001} {\bibfield  {journal} {\bibinfo  {journal} {Phys. Rev. Lett.}\ }\textbf {\bibinfo {volume} {131}},\ \bibinfo {pages} {263001} (\bibinfo {year} {2023})}\BibitemShut {NoStop}%
\bibitem [{\citenamefont {Berry}(1984)}]{MBerry}%
  \BibitemOpen
  \bibfield  {author} {\bibinfo {author} {\bibfnamefont {M.~V.}\ \bibnamefont {Berry}},\ }\href {https://doi.org/10.1098/rspa.1984.0023} {\bibfield  {journal} {\bibinfo  {journal} {Proc. R. Soc. Lond. A}\ }\textbf {\bibinfo {volume} {392}},\ \bibinfo {pages} {45} (\bibinfo {year} {1984})}\BibitemShut {NoStop}%
\bibitem [{\citenamefont {Fukui}\ \emph {et~al.}(2005)\citenamefont {Fukui}, \citenamefont {Hatsugai},\ and\ \citenamefont {Suzuki}}]{hatsu3}%
  \BibitemOpen
  \bibfield  {author} {\bibinfo {author} {\bibfnamefont {T.}~\bibnamefont {Fukui}}, \bibinfo {author} {\bibfnamefont {Y.}~\bibnamefont {Hatsugai}},\ and\ \bibinfo {author} {\bibfnamefont {H.}~\bibnamefont {Suzuki}},\ }\href {https://doi.org/10.1143/JPSJ.74.1674} {\bibfield  {journal} {\bibinfo  {journal} {J. Phys. Soc. Jpn.}\ }\textbf {\bibinfo {volume} {74}},\ \bibinfo {pages} {1674} (\bibinfo {year} {2005})}\BibitemShut {NoStop}%
\bibitem [{\citenamefont {Hatsugai}(2005)}]{hatsu4}%
  \BibitemOpen
  \bibfield  {author} {\bibinfo {author} {\bibfnamefont {Y.}~\bibnamefont {Hatsugai}},\ }\href {https://doi.org/10.1143/JPSJ.74.1374} {\bibfield  {journal} {\bibinfo  {journal} {J. Phys. Soc. Jpn.}\ }\textbf {\bibinfo {volume} {74}},\ \bibinfo {pages} {1374} (\bibinfo {year} {2005})}\BibitemShut {NoStop}%
\bibitem [{\citenamefont {Greschner}\ \emph {et~al.}(2020)\citenamefont {Greschner}, \citenamefont {Mondal},\ and\ \citenamefont {Mishra}}]{edge_mishra}%
  \BibitemOpen
  \bibfield  {author} {\bibinfo {author} {\bibfnamefont {S.}~\bibnamefont {Greschner}}, \bibinfo {author} {\bibfnamefont {S.}~\bibnamefont {Mondal}},\ and\ \bibinfo {author} {\bibfnamefont {T.}~\bibnamefont {Mishra}},\ }\href {https://doi.org/10.1103/PhysRevA.101.053630} {\bibfield  {journal} {\bibinfo  {journal} {Phys. Rev. A}\ }\textbf {\bibinfo {volume} {101}},\ \bibinfo {pages} {053630} (\bibinfo {year} {2020})}\BibitemShut {NoStop}%
\bibitem [{\citenamefont {Mondal}\ \emph {et~al.}(2022)\citenamefont {Mondal}, \citenamefont {Padhan},\ and\ \citenamefont {Mishra}}]{edge_misra_2}%
  \BibitemOpen
  \bibfield  {author} {\bibinfo {author} {\bibfnamefont {S.}~\bibnamefont {Mondal}}, \bibinfo {author} {\bibfnamefont {A.}~\bibnamefont {Padhan}},\ and\ \bibinfo {author} {\bibfnamefont {T.}~\bibnamefont {Mishra}},\ }\href {https://doi.org/10.1103/PhysRevB.106.L201106} {\bibfield  {journal} {\bibinfo  {journal} {Phys. Rev. B}\ }\textbf {\bibinfo {volume} {106}},\ \bibinfo {pages} {L201106} (\bibinfo {year} {2022})}\BibitemShut {NoStop}%
\bibitem [{\citenamefont {Chanda}\ \emph {et~al.}(2022)\citenamefont {Chanda}, \citenamefont {González-Cuadra}, \citenamefont {Lewenstein}, \citenamefont {Tagliacozzo},\ and\ \citenamefont {Zakrzewski}}]{frac_charge_num_Lewsn_2022}%
  \BibitemOpen
  \bibfield  {author} {\bibinfo {author} {\bibfnamefont {T.}~\bibnamefont {Chanda}}, \bibinfo {author} {\bibfnamefont {D.}~\bibnamefont {González-Cuadra}}, \bibinfo {author} {\bibfnamefont {M.}~\bibnamefont {Lewenstein}}, \bibinfo {author} {\bibfnamefont {L.}~\bibnamefont {Tagliacozzo}},\ and\ \bibinfo {author} {\bibfnamefont {J.}~\bibnamefont {Zakrzewski}},\ }\href {https://doi.org/10.21468/SciPostPhys.12.2.076} {\bibfield  {journal} {\bibinfo  {journal} {SciPost Phys.}\ }\textbf {\bibinfo {volume} {12}},\ \bibinfo {pages} {076} (\bibinfo {year} {2022})}\BibitemShut {NoStop}%
\bibitem [{\citenamefont {Jackiw}\ and\ \citenamefont {Rebbi}(1976)}]{jackiw_charge_frac_1976}%
  \BibitemOpen
  \bibfield  {author} {\bibinfo {author} {\bibfnamefont {R.}~\bibnamefont {Jackiw}}\ and\ \bibinfo {author} {\bibfnamefont {C.}~\bibnamefont {Rebbi}},\ }\href {https://doi.org/10.1103/PhysRevD.13.3398} {\bibfield  {journal} {\bibinfo  {journal} {Phys. Rev. D}\ }\textbf {\bibinfo {volume} {13}},\ \bibinfo {pages} {3398} (\bibinfo {year} {1976})}\BibitemShut {NoStop}%
\bibitem [{\citenamefont {Zhou}\ \emph {et~al.}(2023)\citenamefont {Zhou}, \citenamefont {Pan},\ and\ \citenamefont {Jia}}]{zhou_exploring_2023}%
  \BibitemOpen
  \bibfield  {author} {\bibinfo {author} {\bibfnamefont {X.}~\bibnamefont {Zhou}}, \bibinfo {author} {\bibfnamefont {J.-S.}\ \bibnamefont {Pan}},\ and\ \bibinfo {author} {\bibfnamefont {S.}~\bibnamefont {Jia}},\ }\href {https://doi.org/10.1103/PhysRevB.107.054105} {\bibfield  {journal} {\bibinfo  {journal} {Phys. Rev. B}\ }\textbf {\bibinfo {volume} {107}},\ \bibinfo {pages} {054105} (\bibinfo {year} {2023})}\BibitemShut {NoStop}%
\bibitem [{\citenamefont {White}(1992)}]{white_density_1992}%
  \BibitemOpen
  \bibfield  {author} {\bibinfo {author} {\bibfnamefont {S.~R.}\ \bibnamefont {White}},\ }\href {https://doi.org/10.1103/PhysRevLett.69.2863} {\bibfield  {journal} {\bibinfo  {journal} {Phys. Rev. Lett.}\ }\textbf {\bibinfo {volume} {69}},\ \bibinfo {pages} {2863} (\bibinfo {year} {1992})}\BibitemShut {NoStop}%
\bibitem [{\citenamefont {White}(1993)}]{white_density-matrix_1993}%
  \BibitemOpen
  \bibfield  {author} {\bibinfo {author} {\bibfnamefont {S.~R.}\ \bibnamefont {White}},\ }\href {https://doi.org/10.1103/PhysRevB.48.10345} {\bibfield  {journal} {\bibinfo  {journal} {Phys. Rev. B}\ }\textbf {\bibinfo {volume} {48}},\ \bibinfo {pages} {10345} (\bibinfo {year} {1993})}\BibitemShut {NoStop}%
\bibitem [{\citenamefont {Schollw\"ock}(2005)}]{scholwork2005}%
  \BibitemOpen
  \bibfield  {author} {\bibinfo {author} {\bibfnamefont {U.}~\bibnamefont {Schollw\"ock}},\ }\href {https://doi.org/10.1103/RevModPhys.77.259} {\bibfield  {journal} {\bibinfo  {journal} {Rev. Mod. Phys.}\ }\textbf {\bibinfo {volume} {77}},\ \bibinfo {pages} {259} (\bibinfo {year} {2005})}\BibitemShut {NoStop}%
\bibitem [{\citenamefont {Schollwöck}(2011)}]{SCHOLLWOCK2011}%
  \BibitemOpen
  \bibfield  {author} {\bibinfo {author} {\bibfnamefont {U.}~\bibnamefont {Schollwöck}},\ }\href {https://doi.org/https://doi.org/10.1016/j.aop.2010.09.012} {\bibfield  {journal} {\bibinfo  {journal} {Annals of Physics}\ }\textbf {\bibinfo {volume} {326}},\ \bibinfo {pages} {96} (\bibinfo {year} {2011})}\BibitemShut {NoStop}%
\bibitem [{\citenamefont {Hauschild}\ and\ \citenamefont {Pollmann}(2018)}]{hauschild_efficient_2018}%
  \BibitemOpen
  \bibfield  {author} {\bibinfo {author} {\bibfnamefont {J.}~\bibnamefont {Hauschild}}\ and\ \bibinfo {author} {\bibfnamefont {F.}~\bibnamefont {Pollmann}},\ }\href {https://doi.org/10.21468/SciPostPhysLectNotes.5} {\bibfield  {journal} {\bibinfo  {journal} {SciPost Phys. Lect. Notes}\ ,\ \bibinfo {pages} {5}} (\bibinfo {year} {2018})}\BibitemShut {NoStop}%
\bibitem [{\citenamefont {Mishra}\ \emph {et~al.}(2009{\natexlab{b}})\citenamefont {Mishra}, \citenamefont {Pai}, \citenamefont {Ramanan}, \citenamefont {Luthra},\ and\ \citenamefont {Das}}]{mishra_solit_prof_2009}%
  \BibitemOpen
  \bibfield  {author} {\bibinfo {author} {\bibfnamefont {T.}~\bibnamefont {Mishra}}, \bibinfo {author} {\bibfnamefont {R.~V.}\ \bibnamefont {Pai}}, \bibinfo {author} {\bibfnamefont {S.}~\bibnamefont {Ramanan}}, \bibinfo {author} {\bibfnamefont {M.~S.}\ \bibnamefont {Luthra}},\ and\ \bibinfo {author} {\bibfnamefont {B.~P.}\ \bibnamefont {Das}},\ }\href {https://doi.org/10.1103/PhysRevA.80.043614} {\bibfield  {journal} {\bibinfo  {journal} {Phys. Rev. A}\ }\textbf {\bibinfo {volume} {80}},\ \bibinfo {pages} {043614} (\bibinfo {year} {2009}{\natexlab{b}})}\BibitemShut {NoStop}%
\end{thebibliography}%
\end{document}